\documentclass[11pt,a4paper]{article}
\usepackage{jheppub}
\usepackage{graphicx}
\usepackage{pifont}
\usepackage{amsfonts}
\usepackage{amssymb}
\usepackage{amsmath}
\usepackage{hyperref}
\hypersetup{colorlinks=true}
\hypersetup{linkcolor=black}
\hypersetup{citecolor=black}
\hypersetup{urlcolor=black}
\numberwithin{equation}{section}
\usepackage{bbm}
\usepackage{mathtools}
\usepackage{datetime}
\usepackage[table]{xcolor}
\usepackage{draft}
\usepackage{booktabs}

\newcommand{\beq}{\begin{equation}\begin{aligned}}
\newcommand{\eeq}{\end{aligned}\end{equation}}

\def\cC{\mathcal{C}}
\def\cD{\mathcal{D}}
\def\cA{\mathcal{A}}
\def\cF{\mathcal{F}}
\def\bQ{\mathbb{Q}}
\def\bR{\mathbb{R}}
\def\bZ{\mathbb{Z}}
\def\bfK{\mathbf{K}}
\def\bfL{\mathbf{L}}
\def\Gal{\mathrm{Gal}}
\def\Tt{{\sf T}}

\newtheorem{thm}{Theorem}[section]
\newtheorem{lem}[thm]{Lemma}
\newtheorem{prop}[thm]{Proposition}
\newtheorem{cor}[thm]{Corollary}

\usepackage{tikz}
\usetikzlibrary{knots,arrows,decorations.markings,cd,shapes}

\usepackage{adjustbox}

\usepackage{pgfplots}
\usepackage{tikz-3dplot}
\tdplotsetmaincoords{60}{115}
\pgfplotsset{compat=newest}

\title{Higher central charges and topological boundaries in 2+1-dimensional TQFTs}
\author[a]{Justin~Kaidi,\footnote{\tt jkaidi@scgp.stonybrook.edu}}
\author[a]{Zohar~Komargodski,\footnote{\tt zkomargodski@scgp.stonybrook.edu}}
\author[a,b]{Kantaro~Ohmori,\footnote{\tt kant.ohmori@gmail.com}}
\author[a]{Sahand~Seifnashri,\footnote{\tt sahand.seifnashri@stonybrook.edu}}
\author[c,d]{Shu-Heng~Shao\footnote{\tt shuhengshao@gmail.com}}
\affiliation[a]{Simons Center for Geometry and Physics, Stony Brook University,
Stony Brook, NY 11794-3636, USA}
\affiliation[b]{Department of Physics, The University of Tokyo, Bunkyo-ku, Tokyo 113-0033, Japan}
\affiliation[c]{School of Natural Sciences, Institute for Advanced Study, Princeton NJ, USA}
\affiliation[d]{C. N. Yang Institute for Theoretical Physics, Stony Brook University, Stony Brook, NY
11794-3840, USA}
\abstract{A 2+1-dimensional topological quantum field theory (TQFT) may or may not admit topological (gapped) boundary conditions. A famous necessary, but not sufficient, condition for the existence of a topological boundary condition is that the chiral central charge $c_-$ has to vanish. In this paper, we consider conditions associated with ``higher" central charges, which have been introduced recently in the math literature. In terms of these new obstructions, we identify  necessary and sufficient conditions for the existence of a topological boundary in the case of bosonic, Abelian TQFTs, providing an alternative to the identification of a Lagrangian subgroup. 
Our proof relies on general aspects of gauging generalized global symmetries. 
For non-Abelian TQFTs, we give a geometric way of studying topological boundary conditions, and explain certain necessary conditions given again in terms of the higher central charges.
Along the way, we find a curious duality in the partition functions of Abelian TQFTs, which begs for an explanation via the 3d-3d correspondence.
}

\begin{document}
\maketitle

\section{Introduction}

The boundary physics of a Quantum Field Theory is essential in many applications.
A curious phenomenon is that the boundary must sometimes support gapless modes, even though the bulk has an energy gap above its vacuum.
In such cases, the bulk theory is well-described by a Topological Quantum Field Theory (TQFT),   but the boundary does not become topological in the deep IR.
In other words, a TQFT does not always admit a topological boundary condition, and it is natural to ask under what conditions such a  boundary does or does not exist.

Consider a QFT with global symmetry $G$, which could be continuous or discrete. If the symmetry $G$ has an 't Hooft anomaly, then no boundary condition respecting $G$ can exist.\footnote{For instance, take $G=U(1)$ in 3+1 dimensions. To prove that there does not exist a $U(1)$-preserving boundary condition, we assume by contradiction that such a boundary condition does exist. Then we can couple the theory to a background gauge field $A$ and write the gauge variation of the partition function as $\int_{{\cal M}} \alpha F\wedge F$ where $\alpha$ is the gauge variation parameter and $F$ is the field strength. The manifold ${\cal M}$ now has a boundary and hence the variation of the partition function does not make sense since $F\wedge F$ no longer has quantized integrals. If one tries to fix this issue by adding a term $c\int_{{\partial\cal M}} \alpha A\wedge dA$ to the anomalous gauge transformation (with some coefficient $c$), then the Wess-Zumino consistency condition is no longer obeyed since $[\delta_\alpha,\delta_\beta] \log Z = c \int_{\partial{\cal M}} (\alpha d\beta-\beta d\alpha)F$. See \cite{Thorngren:2020yht} for a recent exposition of this topic.} 
Therefore, let us focus on the case in which the symmetry $G$ has no 't Hooft anomalies. 
More specifically, we assume the bulk to be a nontrivial, gapped, $G$-Symmetry Protected Topological (SPT) phase. 
The continuum description of an SPT phase is in terms of a classical field theory of the  $G$ background gauge fields. This is also known as an invertible field theory \cite{Freed:2004yc}.   
In this case $G$-preserving boundary conditions can exist. But if we require that we can couple the theory and its boundary to background $G$-gauge fields in a gauge invariant way,
then the boundary cannot be trivially gapped because the nontrivial SPT phase results in anomaly inflow into the boundary.\footnote{Here we fix the scheme so that the trivial theory on the other side of the boundary has vanishing SPT phase. Then a nontrivial SPT phase in the theory of interest is enough to guarantee a nontrivial boundary.}
In particular, when the SPT corresponds to a perturbative anomaly for a continuous $G$, then the boundary has to be gapless.\footnote{This is because the anomalous Ward identity ensures that the separate $(d-1)$-point function of the currents is nontrivial. }

Another way to think about this situation is in the language of interfaces. Indeed, using the folding trick one can reinterpret an interface between the theories $\cal T$ and ${\cal T}'$ as a boundary condition for the theory ${\cal T}\times \overline{{\cal T'}}$ ($\overline{{\cal T'}}$ stands for the orientation reversal of $\cal T'$).
What we said above therefore implies that a symmetry-preserving interface between $\cal T$ and $\cal T'$ exists only if $\cal T$ and $\cal T'$ have the same anomaly for the symmetry under discussion. 
Furthermore, if $\cal T$ and $\cal T'$ have no anomaly for the symmetry $G$ but are both in some SPT phase, a symmetry-preserving interface can be trivially gapped only if the two SPT phases agree.

In this paper, we consider a more general case in 2+1-dimensions, where the bulk theory is gapped and has nontrivial anyon excitations. The continuum description is in terms of a nontrivial TQFT, rather than a classical field theory. 
The main objective of this paper is to understand when topological boundary conditions for a general 2+1-dimensional TQFT can exist. 
For simplicity, we will only focus on bosonic (non-spin) theories in this paper. 
We will use ``gapped boundary" as a synonym for topological boundary, and  these two terms will be used interchangeably.

The problem of finding topological boundary conditions of a 2+1d TQFT  has a long history in the high energy physics, the condensed matter physics, and the mathematical literature. 
In the case  of abelian TQFTs, this has been discussed in \cite{Kapustin:2010hk} from a field theory point of view. 
Mathematicians have studied this problem in the context of modular tensor category \cite{davydov2013witt,davydov2013structure,Fuchs:2012dt} and of fully extended  field theories \cite{Freed:2020qfy}.
In the condensed matter literature, it has been studied extensively in the context of gapped boundaries of  topological order, e.g.\ in \cite{Haldane:1995xgi,Kitaev:2011dxc,Wang:2012am,2013PhRvX...3b1009L,Barkeshli:2013jaa,Kapustin:2013nva,Hung:2014tba,Lan:2014uaa,Wang:2018edf,Wang:2018qvd,Kong:2019byq}.  
In this paper, we will derive new results and provide a geometric interpretation for some of the known results.

\subsection{The Chiral Central Charge}

We start with a discussion of a non-spin, invertible field theories in 2+1 dimensions without any symmetry. 
Consider the following classical field theory on a three-manifold $M$:
\begin{equation}\label{gCS}
	{\cal L}  ={ \frac{k}{12}\pi }\int_{M} \mathrm{Tr}\left(\omega\wedge d \omega+{2\over 3}\omega\wedge\omega\wedge\omega\right)~,\quad k\in\mathbb{Z}~.
\end{equation}
Here $\omega$ is the  spin connection of $M$. 
This invertible field theory depends on  the metric as well as a choice of the framing of $M$.

To get rid of the framing dependence, we can consider a closely related invertible field theory  given by the $\eta$-invariant with coefficient $k$ (normalized appropriately). 
It is an invertible field theory that is independent of the framing. 
In the condensed matter literature, this is known as Kitaev's $E_8$ phase \cite{kitaev2011toward}. 
Finally, the ratio of Kitaev's $E_8$ phase and \eqref{gCS} is the $(E_8)_1$ Chern-Simons (CS) theory, in the  quantization scheme of  \cite{Witten:1988hf}. 
The latter is topological but depends on the choice of the framing.

On general grounds, we could always add massive degrees of freedom in the UV. 
This amounts to changing the IR field theory by an invertible field theory, such as the  $(E_8)_1$ CS theory. 
This has the effect of shifting the coefficient $k$  by 1.

Next, we would like to discuss boundary conditions of a 2+1-dimensional theory, subject to the freedom of stacking an invertible field theory.

Suppose the 2+1-dimensional theory is  in a gapless phase described by  a CFT.
 The  boundary of  a CFT  must always support a displacement operator, which has a nontrivial scaling dimension and thus renders the boundary non-topological.

The case of primary interest in this paper is when the bulk 2+1-dimensional theory is  a  TQFT.  
Such a TQFT describes the low energy limit of a    gapped  phase with long-range entanglement. 
There are nontrivial vacuum degeneracies on different spatial two-manifolds. 
In this case, one defines the effective coefficient of the gravitational Chern-Simons  term in the infrared~\eqref{gCS} via a contact term in the two-point function of the energy-momentum tensor.
It does not have to be quantized with $k\in \mathbb{Z}$ and it is physically measurable through the so-called  thermal Hall conductance. 
Equivalently, the quantization of a 2+1-dimensional TQFT (such as the Chern-Simons theory) generally leads to a term \eqref{gCS} with non-integral $k$ \cite{Witten:1988hf}.   
The chiral central charge of the bulk TQFT is then defined by 
\ie
c_- = 8k~.
\fe
Since we can stack an invertible field theory on top of our TQFT, $c_-$ is only determined by the anyon data mod $8\mathbb{Z}$.

A famous statement is that if $c_-\notin 8\mathbb{Z}$, then there is no topological boundary condition. 
In this case,  the boundary  theory is not a well-defined 1+1-dimensional theory on its own right.  
Rather, it is a ``relative'' theory, which is not invariant under  diffeomorphisms.  
This has to be captured by the gapless modes living on the boundary. 
Furthermore, these gapless modes cannot be removed by adding ordinary 1+1-dimensional theories on the boundary, which always have properly quantized gravitational anomaly $c_- \in 8\mathbb{Z}$.

An interesting fact is that there are TQFTs with $c_-=0$ mod 8 and yet no topological boundary conditions. Given that the only presently known obstruction due to symmetries and anomalies is the chiral central charge $c_-$, it is puzzling that theories with $c_-=0$ mod 8 can still have protected gapless edge modes. 
One such example is the $U(1)_2\times U(1)_{-4}$ Chern-Simons theory, which  has $c_-=0$, and yet its boundary has to be gapless.

\subsection{Obstructions Beyond Anomalies}

We now give a  summary of our results. 
We start with the  Abelian TQFTs. The basic data of an Abelian TQFT includes a one-form symmetry group $G$, which is generated by the $|G|$ anyons of the theory.

It is  known that an Abelian TQFT admits a gapped boundary if and only if there is a sufficiently large non-anomalous subgroup of $G$, known as a Lagrangian subgroup $L$~\cite{Kapustin:2010hk,Fuchs:2012dt,2013PhRvX...3b1009L,Barkeshli:2013jaa}.  
Gauging the subgroup $L$ of the Abelian TQFT then results in the trivial theory. 
The anyons generating the Lagrangian subgroup can terminate  on this topological boundary. 
This is often referred to as the Abelian anyon condensation~\cite{Bais:2008ni} (see~\cite{Burnell:2017otf} for a review). While this in principle solves the problem of topological boundary conditions for Abelian TQFTs, in practice, the identification of a Lagrangian subgroup of anyons is difficult.  
Our main goal in this paper will be to provide a more readily computable alternative.

We begin by discussing higher obstructions to the existence of topological boundaries, generalizing the chiral central charge $c_-$. 
Recall that the chiral central charge $c_-$, in a preferred choice of scheme, appears as the phase of the partition function of the TQFT on $S^3$.  
One of our main results is to show that a nontrivial phase 
\ie
\xi_M := {Z[M]\over |Z[M]|}
\fe
 of the partition function $Z[M]$ of an Abelian TQFT on a three-manifold $M$ with 
 \ie
 {\mathrm{gcd}}(|H_1(M)|, |G|)=1
 \fe
  is also an obstruction to the existence of a topological boundary condition.  
The proof of this theorem relies on general aspects of gauging a one-form global symmetry.

When the three-manifold $M$ is the lens space $L(n,1)$, 
this phase is known as the \textit{higher central charge} \cite{Ng:2018ddj, ng2020higher}, which we denote by $\xi_n$. The higher central charge admits a simple expression in terms of the spins $\theta(a)$ of the anyons $a$:
\begin{equation}
\xi_n := { \sum_a \theta(a)^n \over| \sum_a \theta(a)^n|}
\end{equation} 
Importantly, we have restricted to the case of  ${\mathrm{gcd}}(n, |G|)=1$  so far.

Our next result is that there is an extension of the higher central charges $\xi_n$ such that ${\mathrm{gcd}}(n, |G|)=1$ is not required, but the more general condition
\begin{equation}
	\mathrm{gcd}\left(n,\frac{2|G|}{\mathrm{gcd}(n,2|G|)}\right)=1 \label{gen.gcd.cond}
\end{equation}
is satisfied. 
We will prove that an Abelian TQFT admits a topological boundary condition if and only if all these extended higher central charges $\xi_n$ satisfying \eqref{gen.gcd.cond} are trivial. 
Since $\xi_{n}=\xi_{n+2|G|}$, there are only finitely many higher central charges to compute. 
These phases  provide  a highly computable alternative to the known condition in terms of Lagrangian subgroups.

Along the way, we will prove that an Abelian TQFT admits  a topological boundary if and only if it is an Abelian Dijkgraaf-Witten theory \cite{Dijkgraaf:1989pz}.\footnote{Note that a Dijkgraaf-Witten theory based on a finite Abelian group can be an Abelian or a non-Abelian TQFT. Here by  Abelian Dijkgraaf-Witten theories we mean those that are Abelian TQFTs.} This gives a complete classification of all possible Abelian topological field theories with gapped boundaries.

As an aside, we report a curious relation for the partition function of Abelian TQFT. 
  Any Abelian TQFT can be described by an Abelian Chern-Simons theory with a  $K$-matrix $\bfK$ \cite{Belov:2005ze} (see also \cite{WALL1963281,https://doi.org/10.1112/blms/4.2.156,Nikulin_1980,Stirling:2008bq}).  
 On the other hand, any three-manifold $M_{\bfL}$ can be obtained by removing  a link from $S^3$ and gluing in a union of solid tori   specified by a \textit{linking matrix} ${\bfL}$. 
 
 Let $Z_{\bfK}[\bfL]$ be the partition function of the Abelian TQFT labeled by $\bfK$ on the three-manifold $M_{\bfL}$ labeled by  $\bfL$. 
 The partition function of an Abelian TQFT does not depend on the detail of the link other than $\mathbf{L}$. We review the basic ingredients in the surgery construction of 3-manifolds in Appendix \ref{app:RT}.

 In Appendix \ref{app.Abelian} we will show that the partition function on $M_{\bfL}$ satisfies the following identity
\begin{equation} Z_{\bfK}[\bfL]= e^{{2\pi i\over8} \, \mathrm{sgn}(\bfK)\mathrm{sgn}(\bfL)} \left|{\det \bfL\over \det \bfK}\right|^{1/2}\overline{Z_{\bfL}[\bfK]}~.\label{eq:ZKLident}\end{equation}
This identity holds as long as the matrices are \textit{coprime} (see Appendix \ref{app.Abelian} for the definition of coprimality of matrices). This symmetry between the Chern-Simons $K$-matrix $\bfK$ and the surgery linking matrix $\bfL$ calls for a geometric interpretation, perhaps in terms of the 3d-3d correspondence \cite{Dimofte:2011ju}.\footnote{On a related topic, \cite{Cho:2020ljj} discusses constructions of 2+1d TQFTs from three-manifold compactifications of the 6d $\mathcal{N}=(2,0)$ theory.}

\subsection{Gapped Boundaries and Lagrangian Algebras}

Next we consider the problem of topological boundaries in non-Abelian TQFTs. For non-Abelian theories, there generically does not exist a one-form symmetry whose gauging leads to the desired topological boundary. In other words, not all gapped boundaries correspond to Abelian anyon condensation. 

As a simple example, take any TQFT $\cal T$ and consider ${\cal T} \times \overline {\cal T}$. This clearly admits a topological boundary since by the folding trick this is the same as the trivial interface between $\cal T$ and itself, which exists tautologically. However, in general,  ${\cal T}\times\overline{\cal T}$ does not have any one-form symmetry, and hence the existence of this gapped boundary cannot always be understood by gauging a Lagrangian subgroup.

Nonetheless, there is an extension of the notion of a Lagrangian subgroup to a \textit{Lagrangian algebra}. The Lagrangian algebra is a certain non-simple anyon with various special properties, such as having quantum dimension coinciding with the total quantum dimension of the TQFT.\footnote{In general, gapless boundary conditions of  a TQFT is not represented by any (possibly non-simple) anyon. The fact that this is the case for  gapped boundary conditions is reminiscent of the situation in 1+1 dimensions, where the conformal boundary conditions are represented by linear combinations of bulk local operators via the Cardy conditions \cite{CARDY1989581}. One can view the constraints that we derive on this non-simple anyon as 2+1 dimensional counterparts of the Cardy conditions.} As we will review in Section \ref{sec:TV}, there is a one-to-one correspondence between Lagrangian algebras and topological boundary conditions in unitary 2+1d TQFTs \cite{davydov2013witt,Fuchs:2012dt}.

We will provide a geometric picture and construction of the Lagrangian algebra from a topological boundary. This description pictorially trivializes the defining properties of a Lagrangian algebra.
Inserting a fine mesh of this algebra object allows us to formalize the notion of ``condensing non-Abelian anyons." 
We will see that the condensation of these anyons  again leads to a gapped boundary.  

For the particular case of ${\cal T}\times\overline{\cal T}$, if we label the anyons by $a_i$ and those of $\overline{\cal T}$ by $\tilde a_i$ (where $\tilde a_i$ is obtained by time reversal from $a_i$) then the Lagrangian algebra is the following (non-simple) anyon:
\ie
\cA = \bigoplus_i a_i \otimes \tilde a_i~.
\fe

The non-Abelian anyon condensation in many ways appears as a natural generalization of ``gauging" these non-Abelian anyons.\footnote{We thank F. Burnell, T. Devakul, P. Gorantla, H. T. Lam, and N. Seiberg for many discussions on this point.} 
However, while there is a well-defined mathematical procedure, there is no clear physical understanding of what it means to ``gauge" such an algebra anyon other than the insertion of a mesh. In particular, there is no clear understanding in terms of a sum over some gauge fields.\footnote{The generalized notion of gauging in terms of inserting a fine mesh of (non-invertible) topological lines in 1+1d QFTs was introduced by \cite{Frohlich:2009gb, Carqueville:2012dk,Brunner:2014lua} and reviewed in \cite{Bhardwaj:2017xup}. See \cite{Gaiotto:2019xmp} for a discussion in the context of category theory. }

The existence of Lagrangian algebras is studied in the context of modular tensor categories by mathematicians. By the correspondence between Lagrangian algebras and topological boundaries, this is closely related to the problems we are interested in this work. The basic idea is that one can define an equivalence relation between 2+1d TQFTs whenever there exist a topological interface between them. More precisely, we say topological theories ${\cal T}_1$ and ${\cal T}_2$ are \emph{Witt equivalent}, if the theory ${\cal T}_1 \times \overline{\cal T}_2$ has a gapped boundary (or equivalently, has a Lagrangian algebra)~\cite{davydov2013witt,davydov2013structure}. These equivalence classes form an Abelian group called the Witt group. The group structure is given by taking tensor product of two TQFTs. More precisely, the group multiplication is defined by $[{\cal T}_1]\cdot[{\cal T}_2] := [{\cal T}_1 \times {\cal T}_2]$, where $[{\cal T}]$ denotes the Witt class of the theory $\cal T$. The group inverse is given by $[{\cal T}]^{-1}=[\overline{\cal T}]$, since ${\cal T}\times \overline{\cal T}$ has a gapped boundary and hence is trivial in the Witt group. 

Having introduced the Witt group, we will review the theorem~\cite[Theorem 4.4]{Ng:2018ddj} that the phase of lens space partition functions $\xi_n$ with $\mathrm{gcd}(n,N_\mathrm{FS})=1$, are invariants of the Witt group. In other words, these phases are obstructions to the existence of a gapped boundary for non-Abelian theories. Here $N_\mathrm{FS}$ is the \textit{Frobenius-Schur exponent}, i.e.\ the smallest integer such that $\theta(a)^{N_\mathrm{FS}}=1$ for  of all anyons $a$. 
This shows that even in the non-Abelian case there are important obstructions beyond the chiral central charge.

Along the way we will uncover a few elementary but interesting properties of the lens space partition functions in theories with a gapped boundary. These properties will be derived by making use of Galois theory.

For the convenience of the reader, we summarize various equivalent properties of TQFTs in  Table \ref{table}. 
The details   will be discussed    in the main body of the paper. 
See, in particular, Section \ref{sec:completeObs} and \ref{sec:TV} and references therein. 
 In the lower right corner, although there is no analogous condition like the vanishing of $\xi_n$ at present, we will discuss necessary conditions for the existence of a topological boundary.

\begin{table}[]
	\renewcommand{\arraystretch}{1.5} 
	\centering
	\begin{tabular}{|c|c|}
		\hline
		\textbf{Abelian TQFTs }                  & \textbf{Non-Abelian TQFTs}                    
		\\ \hline 
		$\exists$~	Topological boundary condition  & $\exists$~Topological boundary condition       \\  \hline
		$\exists$~	Lagrangian subgroup             & $\exists$~Lagrangian algebra                   \\ \hline
		Abelian Dijkgraaf-Witten theory &~~ Turaev-Viro theory / Drinfeld center~~ \\ \hline 
		$~~\xi_n = { \sum_a \theta(a)^n \over| \sum_a \theta(a)^n|}=1$ for ${\rm gcd}\left( n, {2 |G| \over {\rm gcd}(n,2|G|)}\right) =1~~$  & ?                                    \\ [1.2ex] \hline
	\end{tabular}
	\caption{We summarize various sufficient and necessary conditions for the existence of topological boundary conditions, both in the case of Abelian and non-Abelian TQFTs. The conditions in each column are equivalent to each other.   In the first column, $|G|$ denotes the number of simple anyons, and $\xi_n$ is the higher central charge (i.e., the phase of the $L(n,1)$ lens space partition function).  }
	\label{table}
\end{table}

\subsection*{Organization}
The rest of this paper will be organized as follows. In Section \ref{sec:gapAbelian} we will focus on Abelian theories, beginning with a review of the known results about Lagrangian subgroups in Section \ref{sec:Lsubgroup}. Then in Section \ref{sec:main} will discuss the new class of obstructions $\xi_M$ labelled by 3-manifolds $M$, including the higher central charges. We then extend the higher central charges to a complete set of obstructions in Section \ref{sec:completeObs}, giving an alternative set of necessary and sufficient conditions beyond the usual Lagrangian subgroups. We end the Abelian discussion with Section \ref{sec:anotherpov}, which gives an alternative viewpoint on these obstructions. This alternative viewpoint partly generalizes to the non-Abelian case and allows to establish several general result about the properties of $\xi_M$ in Abelian theories.  
For instance, the phases $\xi_M$ are always 8th roots of unity. 

We then move on to a discussion of the non-Abelian case in Section \ref{sec:nonAbel}, beginning with a detailed geometric definition of Lagrangian algebra anyons in Section \ref{sec:Lagalg}. Utilizing various geometric considerations, we find many properties that Lagrangian algebra anyons must satisfy.

In Section \ref{sec:gauging} we discuss a generalized notion of gauging of the Lagrangian algebra. 
Section \ref{sec:TV} reviews various sufficient and necessary conditions for the existence of a topological boundary conditions for general non-Abelian TQFTs. 
Section \ref{sec:nonAbhigherc} discusses the non-Abelian analogs of the higher central charges. Using Galois theory we find several general properties of the higher central charges.

Finally, some background information is collected in the Appendices. Appendix \ref{app:3d.TFT} gives a brief review of 2+1d TQFT. Appendix \ref{app:RT} reviews  the surgery construction of TQFT partition functions. Appendix \ref{app.Abelian} obtains various results regarding the partition function of Abelian Chern-Simons theories labelled by the matrix $\bfK$ on manifolds with linking matrix $\bfL$, including the identity (\ref{eq:ZKLident}) given above.

\section{Abelian TQFTs}\label{sec:gapAbelian}

In this section we discuss the question of topological boundary conditions in Abelian TQFTs.  
Along the way, we will phrase some of the results in the literature in terms of gauging higher-form symmetries.
Throughout this section, we will make use of various basic properties of 2+1d TQFT reviewed in Appendix \ref{app:3d.TFT}.

\subsection{Lagrangian Subgroups}\label{sec:Lsubgroup}

A \textit{Lagrangian subgroup} $L$ \cite{2009arXiv0909.3140E} of an Abelian TQFT  is a subgroup of the bosons (i.e. anyons with spin $\theta(a)=1$) such that $|L|^2=|G|$, where $G$ is the Abelian group obtained from the fusion of anyons. 
It follows from this definition that\footnote{The first property is obvious from the expression of the braiding phase in terms of the spins. The second property can be shown as follows. Since elements in $L$ braid trivially with each other, the braiding phase $B(a,b)$ induces a homomorphism from $L \times G/L$ to $U(1)$, which can be alternatively viewed as a homomorphism $h : L \to \widehat{G/L}$, where $\widehat H=  \text{Hom}(H,U(1))$ is the Pontryagin dual of $H$. 
Since the braiding is non-degenerate, $h$ must be injective. Moreover, the Lagrangian condition $|L|^2 =|G|$ implies that $h$ must be surjective as well. The homomorphism $h$ being surjective then implies that the homomorphism $h' :G/L \to \widehat{L}$ is injective, which proves the second property.}
\begin{enumerate}
\item Every two lines in $L$ have trivial braiding, i.e. $B(a,b) = {\theta(a\times b)\over \theta(a)\theta(b)}=1$ for all $a,b\in L$.
\item Any line that is not in $L$ has nontrivial braiding with at least one line in $L$. \label{property.2}
\end{enumerate}
Note that the choice of the Lagrangian subgroup for a given Abelian TQFT is generically not unique.

In  \cite{Kapustin:2010hk,Fuchs:2012dt,2013PhRvX...3b1009L,Barkeshli:2013jaa} the following result was shown,
\begin{thm}
\label{thm:Lagsub}
A $c_-=0$ Abelian bosonic TQFT has a gapped boundary if and only if there exists a Lagrangian subgroup $L$.
\end{thm}
\noindent
We now provide an interpretation of this result in terms of gauging one-form symmetries.

\subsubsection*{Gauging One-Form Symmetries and Dijkgraaf-Witten Theories}

The anyons of an Abelian TQFT generate a one-form symmetry group $G$, and the 't Hooft anomalies of $G$ are captured by the spins $\theta(a)$ of the anyons \cite{Gaiotto:2014kfa,Gomis:2017ixy,Hsin:2018vcg}. 
In particular, one-form symmetries generated by bosonic lines   are non-anomalous and can be gauged.
(Note however that the set of the bosons in an Abelian TQFT is generally not closed under fusion, and hence does not form a group). In this language, Theorem \ref{thm:Lagsub} can  be restated as:
\begin{itemize}
\item A $c_-=0$ Abelian bosonic TQFT has a gapped boundary if and only if there is a non-anomalous subgroup $L$ of the one-form symmetry group $G$ such that $|L|^2=|G|$.
\end{itemize}
\noindent
Another statement, more within the framework of the low-energy TQFT which allows us to define $c_-$ only mod 8, is

\begin{itemize}
\item An Abelian bosonic TQFT has a gapped boundary after stacking with appropriate copies of the $(E_8)_1$ Chern-Simons theory if and only if there is a non-anomalous subgroup $L$ of the one-form symmetry group $G$ such that $|L|^2=|G|$.
\end{itemize}

As explained in the introduction, stacking copies of the $(E_8)_1$ Chern-Simons theory does not change  anyon date of  the TQFT. All it does is to shift the gravitational Chern-Simons coefficient $k$ from~\eqref{gCS} by an integer. Equivalently, stacking copies of the $(E_8)_1$ Chern-Simons theory adds edge modes with $c_-\in 8\mathbb{Z}$.

Note that in 2+1d, every time we gauge a non-anomalous discrete one-form  symmetry $L$, we gain  a quantum zero-form symmetry $\widehat L$  in the gauged theory, where $\widehat L=\text{Hom}(L,\bR/\bZ)$ is the Pontryagin dual of $L$ \cite{Gaiotto:2014kfa,Tachikawa:2017gyf}. 
This is the 2+1d version of the quantum symmetry of the orbifold theory in 1+1d \cite{VAFA1986592}.

Let us explain this point further. 
First recall that coupling the theory to a background gauge field is equivalent to the insertion of a certain network of $L$ symmetry lines into the spacetime manifold $M$. For a discrete and anomaly-free $L$, the insertions only depend on the homology class of the network, and hence by Poincar\'e duality the network is characterized by a cocycle $a\in H^2(M,L)$. Gauging the symmetry is equivalent to summing over gauge fields and hence making $a$ dynamical. Now we can introduce a background gauge field $\hat{a} \in H^1(M,\widehat{L})$, and add the term $\langle \hat{a} , a \rangle$ to the action. The gauge field  $\hat{a} \in H^1(M,\widehat{L})$ is a gauge field for the quantum $\widehat{L}$ zero-form symmetry. The partition function of the gauged theory becomes  
\begin{equation}
	Z'[M,\hat{a}] =  { |H^0(M,L) |\over |H^1(M,L)| } \sum_{a \in H^2(M,L)} e^{2\pi i \, \langle \hat{a} , a \rangle} \, Z[M,a]~.
\end{equation}
Here $\langle.,.\rangle: H^1(M,\widehat{L}) \times H^2(M,L) \rightarrow \bR/\bZ$ is the intersection pairing on cohomology. $ Z[M,a]$ stands for the partition function on manifold $M$ coupled to a background gauge field $a$.
An important property of the quantum  $\widehat L$ zero-form symmetry is that it can be gauged to retrieve the original theory, since
\begin{equation}
	{ 1\over |H^0(M,\widehat{L})| } \sum_{\hat{a} \in H^1(M,\widehat{L})} Z'[M,\hat{a}] = Z[M,0]~.
\end{equation}

Now let us consider an Abelian bosonic TQFT  with a Lagrangian subgroup $L$.
We can gauge this non-anomalous one-form symmetry subgroup $L$ to obtain another TQFT. 
Following the three-step process of gauging a one-form symmetry outlined in \cite{Moore:1988ss,Moore:1989yh,Hsin:2018vcg}, and using the properties of $L$ described at the beginning of this section, one finds that the gauged TQFT is trivial. 
But if this is the case, then as was just reviewed the original TQFT can be retrieved from the trivial TQFT by gauging an $\widehat L$ zero-form symmetry, possibly with a nontrivial  action for the $\widehat L$ gauge fields.  
In other words, the   TQFT  is  a finite group gauge theory $\widehat L$, possibly with a nontrivial  Dijkgraaf-Witten twist \cite{Dijkgraaf:1989pz}.\footnote{We can find the particular element of $H^3(\widehat L, U(1))$  specifying the Dijkgraaf-Witten twist as follows. By restricting the braiding phase $B:G \times G \to U(1)$ to $G/L \times L$, we obtain a non-degenerate pairing between $G/L$ and $L$, which provides an isomorphism between $G/L$ and $\widehat L$ \cite{2009arXiv0909.3140E}. Since $G$ can be viewed an extension of $G/L \simeq \widehat L$ by $L$, this determines a class in $H^2(\widehat L,L)$.  This class can be further mapped to a 3-cocycle in $H^3(\widehat L, U(1))$ which is  the Dijkgraaf-Witten twist \cite{Kapustin:2010hk}.}

Since conversely every Dijkgraaf-Witten gauge theory also admits a gapped boundary condition (i.e. the Dirichlet boundary condition) we are led the following result:\footnote{See, for example, \cite{Gaiotto:2020iye} for a recent discussions on boundary  conditions of the Dijkgraaf-Witten gauge theory.}
\begin{cor}
\label{cor:DW}
 An Abelian bosonic TQFT has a gapped boundary if and only if it is an Abelian Dijkgraaf-Witten gauge theory.
\end{cor}
\noindent
This statement also follows from the discussion in \cite{Kapustin:2010hk,Fuchs:2012dt}.

\subsubsection*{Example: $\mathbb{Z}_2$ Gauge Theory}

Let us consider  $\mathbb{Z}_2$ gauge theory as an example. The Lagrangian is given by \cite{Maldacena:2001ss,Banks:2010zn,Kapustin:2014gua},
\ie\label{ZtwoDW}
{\cal L} ={2\over2\pi} a^{(1)}db^{(1)}\,,
\fe
and the theory admits four anyons,
\ie \label{Lines}
1\,,~~~~e\equiv  e^{ i \oint a^{(1)}} \,,~~~~m  \equiv e^{i \oint b^{(1)}} \,,~~~~ f \equiv e^{i \oint a ^{(1)}+ i\oint b^{(1)}}\,.
\fe
Their spins are $\theta(1) = \theta(e) =\theta(m) =1$ and $\theta(f)=-1$. 
They obey the $\mathbb{Z}_2\times \mathbb{Z}_2$ fusion rules,
\ie
&e\times e= m\times m = f\times f =1\,,~~~~~ e\times m = m\times e = f\,,\\
& m\times f = f\times m =e \,,~~~~~~~~~~~~~~~~\, f\times e= e\times f = m \,.
\fe

The subset of bosonic anyons is $\{1,e,m\}$, but this is not closed under fusion and hence does not form a subgroup. Instead, we have the following two bosonic subgroups, 
\ie
&\mathbb{Z}_2^{(e)} = \{ 1, e\} \,,~~~~~\mathbb{Z}_2^{(m)} = \{ 1,m\}\,.
\fe
Since $2^2 = 4$, each of these is a Lagrangian subgroup, and there are consequently two corresponding choices for gapped boundaries conditions.
Indeed, gauging either one of the two Lagrangian subgroups (but not both) gives the trivial theory. 
For example, gauging $\mathbb{Z}_2^{(e)}$ gives
\ie
{2\over 2\pi} a^{(1)} ( db ^{(1)}- c^{(2)}) + {2\over 2\pi } c^{(2)} d\phi ^{(0)}
\fe
where $c^{(2)}$ is a two-form gauge field and $\phi^{(0)}$ is a $2\pi$-periodic compact scalar. 
The latter enforces the condition that $c^{(2)}$ is a $\mathbb{Z}_2$-valued two-form gauge field. 
The gauge transformations are
\ie
&a^{(1)}\sim a^{(1)} +d\alpha^{(0)}\,,~~~~b^{(1)}\sim b^{(1)}+d\beta^{(0)}+\gamma^{(1)}\,,\\
&\phi^{(0)} \sim \phi^{(0)} + \alpha^{(0)}\,,~~~~\,\,c^{(2)} \sim c^{(2)} + d\gamma^{(1)}\,.
\fe
In terms of the gauge-invariant combination $\bar a^{(1)} := a^{(1)}  - d\phi^{(0)}$, we can rewrite the gauged Lagrangian  as ${2\over 2\pi} \bar a^{(1)}( db^{(1)}- c^{(2)} )$ modulo total derivatives. After integrating out $\bar a^{(1)}$, this indeed becomes a trivial TQFT.

The gapped boundary separates this trivial phase from the $\mathbb{Z}_2$ gauge theory phase. To demonstrate that the boundary is empty let us analyze the theory~\eqref{ZtwoDW} on a half-plane $y\geq0$. Choosing the gauge $a_t=b_t=0$, we are restricted by the Gauss law to satisfy that the $xy$ components of both field strengths vanish, leaving us with flat connections. If we impose at the boundary $a_{||}=0$ (resp. $b_{||}=0$) then we do not have to restrict the $b$ (resp. $a$) gauge transformations at the boundary, and hence the $b$ (resp. $a$) flat connections can be removed everywhere on the disk leaving us with an empty theory. 

Each of these boundary conditions breaks the $\mathbb{Z}_2\times\mathbb{Z}_2$ one-form symmetry to $\mathbb{Z}_2$. For instance, if we impose $a_{||}=0$ then the $\mathbb{Z}_2^{(m)}$ one-form symmetry is broken and the Wilson line $e$ from~\eqref{Lines} can end (condense) on the boundary.

\subsection{Obstructions from 3-Manifold Invariants}\label{sec:main}

We have seen that the existence of a gapped boundary in an Abelian TQFT is tantamount to the existence of a Lagrangian subgroup. In practice though, it is not always straightforward to check if such a subgroup exists. In this subsection, we will provide a new set of obstructions which have the virtue of being highly computable, though with the disadvantage that they will provide only \textit{necessary}, not sufficient, conditions to the existence of a gapped boundary. In the following subsection, we will generalize these obstructions to ones yielding necessary and sufficient conditions.

To begin, let $Z_{\cal T}[M]$ be the partition function of a $c_-=0$  Abelian bosonic TQFT $\cal T$  on a closed, oriented, connected three-manifold $M$.  
Since $c_-=0$, we can choose a scheme in which the partition function is topological and independent of the choice of the framing. In this scheme $Z_{\cal T}[M]$ coincides with the Reshetikhin-Turaev invariant of three-manifolds, which is reviewed in Appendix~\ref{app:RT}.

As we have seen, the Abelian TQFT $\cal T$ has a gapped boundary if and only if there exists a non-anomalous one-form symmetry subgroup $L$ with $|L|^2=|G|$, where $G$ is the Abelian group of anyons. 
Furthermore, the TQFT obtained by gauging $L$ is trivial. 
In equations, this means that
\ie\label{gaugeL}
1 = { |H^0(M,L) |\over |H^1(M,L)| } \sum_{a\in H^2 ( M,L) } Z_{\cal T}[M,a]
\fe
where $a$ is a two-form gauge field of the one-form symmetry $L$, and $Z_{\cal T}[M,a]$ is the partition function  coupled to the gauge field $a$.

Consider a three-manifold $M$ with the following property:
\ie\label{gcdHG}
\text{gcd}( |H_1(M) | , |G|) =1\,.
\fe
With this condition we also demand that $H_1(M) = H_1(M,\mathbb{Z})$ has a finite order.
For any Lagrangian subgroup $L$, this condition is equivalent to 
\ie\label{gcdHL}
\text{gcd}( |H_1(M) | , |L|) =1
\fe
since $|L|^2=|G|$. Next, \eqref{gcdHL}  is equivalent to
\ie
H_1( M, L) = 0\,.
\fe
as follows from $H_1(M,L) = H_1(M) \otimes L$ and the fact that $\mathbb{Z}_n \otimes \mathbb{Z}_m = \mathbb{Z}_{\text{gcd}(n,m)}$.\footnote{Here $\otimes$ is the tensor product of Abelian groups, not to be confused with the direct product $\times$. See \cite{hatcher2002algebraic} for a definition.}
Poincare duality further implies that
\ie
H^2( M,L) \cong H_1(M,L) =0\,,
\fe
and using the universal coefficient theorem, we also have $H^1(M,L)\cong\mathrm{Hom}(H_1(M),L)=0$.

Altogether, we conclude that on a manifold satisfying \eqref{gcdHG}, one has $H^2( M,L) =0$ and hence there are no nontrivial two-form gauge fields of the one-form symmetry $L$. 
It follows that gauging $L$ is completely trivial on such manifolds, and so \eqref{gaugeL} reduces to
\ie
Z_{\cal T}[M]=  {1\over |L|}\,,~~~~~~\text{if}~~~\text{gcd}( |H_1(M)|  , |G|)=1\,,
\fe
where we have used $H^0(M,L) = L$ and $H^2( M,L) =0$. This gives rise to a new set of obstructions: namely, the phase of $Z_{\cal T}[M]$ on manifolds satisfying  \eqref{gcdHG} is an obstruction to the existence of a gapped boundary. We summarize this by means of the following theorem,

{\thm A $c_-=0$ Abelian bosonic TQFT $\cal T$ has a gapped boundary only if  $Z_{\cal T}[M] >0$ on every closed oriented three-manifold $M$ with $\text{gcd}( |H_1(M)|  , |G|)=1$. \label{theorem}}
\newline

Let $M=\overline L(n,1)$ be the orientation reversal of the $n$-th lens space.\footnote{We choose to work with $ \overline L(n,1)$ as opposed to $L(n,1)$ in order to avoid an inconvenient sign in $n$ below.} In this case $H_1(M) = \mathbb{Z}_n$.
Using the surgery presentation of lens spaces (reviewed in Appendix  \ref{app:RT}), the  partition function of a $c_-=0$ Abelian bosonic TQFT $\cal T$  on $\overline L(n,1)$ is given by
\ie\label{LensP}
Z_{\cal T}[\overline L(n,1)] = {1\over |G|} \sum_a \theta(a)^n  \,.
\fe
The phases of these partition functions are then given by
\ie
\label{eq:highercentdef}
\xi_n := { \sum_a \theta(a)^n \over| \sum_a \theta(a)^n|}
\fe
These quantities have made a previous appearance in the math literature, where they went under the name of \textit{higher central charges}  \cite{Ng:2018ddj,ng2020higher}. By using various techniques from Galois theory, the authors of those references were able to show that the higher central charges with $\mathrm{gcd}(n, |G|)=1$ are indeed obstructions to having a Lagrangian subgroup.  Now we have seen that the higher central charges are just special cases of the more general obstructions of Theorem \ref{theorem}, which can be obtained from any three-manifold with $\text{gcd}( |H_1(M)|  , |G|)=1$.

\subsubsection*{Example: $U(1)_{2N_1}\times U(1)_{-2N_2}$}

One of the simplest bosonic Abelian TQFTs is $U(1)_{\pm 2N}$ with $N\in \mathbb{N}$.  
 The spins of the anyons are
	 \ie
	 \theta(s) = \exp\left[2\pi i {s^2 \over 4N}\right]\,,~~~~~s=0,1,\cdots, 2N-1\,.
	 \fe
	These anyons generate a one-form symmetry group $G=\mathbb{Z}_{2N}$.  The order of the $T$-matrix $T_{ss'} =\delta_{ss'}\theta(s)$ is $4N$. 
	
Because this theory has chiral central charge $c_- = \pm1$ mod 8, it cannot admit a gapped boundary.  
However, we can instead consider $U(1)_{2N_1}\times U(1)_{-2N_2}$ with $N_1,N_2\in \mathbb{N}$, which has vanishing chiral central charge $c_-=0$ mod 8.  In this case the total one-form symmetry is $G=\mathbb{Z}_{2N_1}\times\mathbb{Z}_{2N_2}$ and the total number of anyons is $|G| = 4 N_1 N_2$.

As reviewed above, a bosonic Abelian TQFT admits a gapped boundary if and only if it has a Lagrangian subgroup. 
 The $K$-matrix of $U(1)_{2N_1}\times U(1)_{-2N_2}$ is 
	 \ie
	 K = \left(\begin{array}{cc}2N_1 & 0 \\0 & -2N_2\end{array}\right)
	 \fe 
and in this case it turns out that the existence of a Lagrangian subgroup is equivalent to finding a two-dimensional integer vector $\Lambda$ such that $\Lambda^T K\Lambda =0$ (see, for example, \cite{2013PhRvX...3b1009L}).
	 We thus conclude that $U(1)_{2N_1}\times U(1)_{-2N_2}$ has a gapped boundary if and only if
	 \ie\label{perfectsq}
	 \sqrt{N_1 N_2} \in \mathbb{N}\,.
	 \fe
Note that the only if direction in~\eqref{perfectsq} is trivial. There is a Lagrangian subgroup of anyons only if the total number of anyons is a square, but $4N_1N_2$ is a square if and only if $N_1N_2$ is a square.
	 
Below we will rephrase this condition in terms of the higher central charges. 
Let us compute the higher central charges of $U(1)_{2N_1}\times U(1)_{-2N_2}$ using (\ref{eq:highercentdef}).  
We begin with the generalized Gauss sum,
	 \ie\label{U1tau}
\sum_{s=0}^{2N-1} \theta(s)^n = \sum_{s=0}^{2N-1} \exp\left[2\pi i n  {s^2 \over 4N}\right] = {1+i\over2} \varepsilon_n^{-1} \sqrt{4N}  \left( {4N\over n}\right)\quad\quad \text{when $\mathrm{gcd}(n,4N)=1\,$,}
	 \fe
where
\ie
\varepsilon_n= \begin{cases}
1\,,~~~~~\text{if}~~~~n=1~\text{mod}~4\\
i\,,~~~~~\text{if}~~~~n=3~\text{mod}~4\\
\end{cases}
\fe
and $\left({a\over c}\right)$ is the Jacobi symbol. (See, for example, Appendix B of \cite{Hsin:2018vcg} for the definition.)
The higher central charges of $U(1)_{2N_1}\times U(1)_{-2N_2}$ can then be easily obtained:
\ie\label{U1U1xi}
\xi_n ( U(1)_{2N_1}\times U(1)_{-2N_2} ) = \left( {N_1 N_2\over n } \right)   \,,
\fe
where we have used  $\left( {ab\over n} \right) =\left( {a\over n} \right) \left( {b\over n} \right)$. 
	 Recall that the higher central charges $\xi_n$ are defined only for $\text{gcd}(n, |G|)=\text{gcd}(n, 4N_1 N_2)=1$, for which the Jacobi symbol $\left( {N_i\over n}\right)$ is always $\pm1$.

To show equivalence with the Lagrangian subgroup condition, we use the following result in number theory\footnote{This proposition was proved, for example, in \cite{hall1933quadratic}.}
{\prop
$N$ is a perfect square if and only if
\ie\label{prop}
\left( {N\over n} \right)=1\,,~~~~\forall ~~\text{odd prime $n$ such that}~~\text{gcd}(n, N)=1\,.
\fe}
\noindent
It then follows that the condition \eqref{perfectsq} is equivalent to the triviality of all the higher central charges \eqref{U1U1xi}. 
We have therefore shown that the Abelian TQFT $U(1)_{2N_1}\times U(1)_{-2N_2}$ admits a gapped boundary if and only if all of its higher central charges $\xi_n$ are trivial. 

\subsubsection*{Abelian TQFTs with Vanishing Higher Central Charges}

With this initial success, one might be tempted to suppose that all Abelian TQFTs with trivial higher central charges admit a gapped boundary. But as we will now see, this assumption is incorrect. 
There are Abelian TQFTs  with trivial $\xi_n$ for all $\text{gcd}(n, |G|)=1$, and yet they do not admit any gapped boundary \cite{ng2020higher}.

Let us begin by  noting that for every prime number $p$, there is a unique TQFT $A_p$ with the following properties: 
\begin{itemize}
	\item Fusion rules $G=\bZ_p \times \bZ_p$.
	\item No non-anomalous one-form symmetry.
	\item $\xi_n=-1$ for all $n$ with $\text{gcd}(n,p)=1$.
\end{itemize}

For example, for $p=2$ the TQFT in question is $A_2=Spin(8)_1 \text{ CS theory}$.
For $p$ odd, the theory can be decomposed as $A_p = {\cal A}^{p,2} \times {\cal A}^{p,-2m}$, where $m$ is an integer such that $\left( {m \over p} \right)=-1$.
Here we are using the notation introduced in \cite{Hsin:2018vcg}, where ${\cal A}^{p,m}$ are defined to be the minimal TQFTs with $\bZ_p$ fusion rule and spins $\theta(a) = \exp(2\pi i \frac{m a^2}{2p})$ for $a\in\bZ_p$. 

Since $A_p$ has no non-anomalous one-form symmetry, it follows that $A_p \times A_q$ for $p\neq q$ also has no non-anomalous one-form symmetry. The theory thus  cannot admit a gapped boundary. But because the higher central charges of each factor theory are $-1$, the higher central charges of the product theory are trivial (in particular, the chiral central charge vanishes mod 8). Hence the theories $A_p \times A_q$ for $p\neq q$ provide examples of theories which do not admit a gapped boundary, but have trivial higher central charges with  $\text{gcd}(n, |G|)=1$.\footnote{In fact these theories and their products generate all the elements in the Witt group of Abelian theories with trivial higher central charges~\cite{ng2020higher}.}

\subsection{Complete Obstructions to Gapped Boundaries}\label{sec:completeObs}

For an Abelian TQFT with one-form symmetry of order $|G|$, we have seen that a necessary condition for the presence of a gapped boundary is the triviality of the higher central charges $\xi_n$, as defined in~\eqref{eq:highercentdef}, for all $n$ such that ${\rm gcd}(n, |G|)=1$. We now show that expanding the range of $n$, we can also obtain a necessary and sufficient condition. In particular, we will show the following:

{\thm An Abelian TQFT $\cal T$ with Frobenius-Schur exponent $N_\mathrm{FS}$ admits a gapped boundary if and only if $\xi_n({\cal T})=1$ for all $n$ such that ${\rm gcd}\left( n, {N_\mathrm{FS} \over {\rm gcd}(n,N_\mathrm{FS})}\right) =1$.  \label{thm:compobs}}
\newline\newline
Recall that the Frobenius-Schur exponent  $N_\mathrm{FS}$ is defined as  the smallest integer such that $\theta(a)^{N_\mathrm{FS}}=1$ for  of all anyons $a$. 
Three remarks are in order: 
\begin{itemize}
	\item Note that the condition on $n$ can be restated as follows. Take the prime factorization $N_\mathrm{FS} = N_1 N_2 \dots N_k$ of $N_\mathrm{FS}$, where $N_i={p_i}^{\alpha_i}$ for distinct prime numbers $p_1, \dots, p_k$. Then for each $i$, either $N_i$ divides $n$ or $ {\rm gcd} (n,N_i) = 1$.\footnote{In other words, for any common prime factor $p$ of $n$ and $N_\mathrm{FS}$, the exponent of $p$ in $n$ must be greater than or equal to that in $N_\mathrm{FS}$.} 
	\item Since $\xi_n=\xi_{n+N_\mathrm{FS}}$, it suffices to check different $n$ modulo $N_\mathrm{FS}$.
	\item For any positive integer $k$, we can replace $N_\mathrm{FS}$ by $kN_\mathrm{FS}$ in Theorem \ref{thm:compobs}. In other words, if we scan over all $0<n<kN_\mathrm{FS}$ such that ${\rm gcd}\left( n, {k N_\mathrm{FS} / {\rm gcd}(n,k N_\mathrm{FS})}\right) =1$, then $n$ mod $N_\mathrm{FS}$ scans over exactly all solutions of ${\rm gcd}\left( n, {N_\mathrm{FS} / {\rm gcd}(n,N_\mathrm{FS})}\right) =1$.
	 In particular, we can replace $N_\mathrm{FS}$ by $2|G|$  since $N_\mathrm{FS}$ divides $2|G|$ (which follows 	   from \eqref{spin.k.matrix} below).
\end{itemize}

In order to prove Theorem \ref{thm:compobs}, we will use some simple facts about the factorization of Abelian TQFTs. We begin with the following, 

{\lem An Abelian TQFT $\cal T$ with Frobenius-Schur exponent $N_\mathrm{FS}$ admits a factorization 
\bea
\label{eq:AbTFTfact}
{\cal T} ={\cal T}_{p_1} \times{\cal T}_{p_2} \times \dots \times{\cal T}_{p_k}
\eea 
where ${\cal T}_{p_i}$ are TQFTs labelled by distinct primes $p_i$, such that the number of anyons in ${\cal T}_{p_i}$ is a positive integer power of $p_i$. If we denote the Frobenius-Schur exponents of ${\cal T}_{p_i}$ by $N_i$, then $N_\mathrm{FS}= N_1 N_2 \dots N_k$. \label{lem:factlem}
}
\newline\newline
At the level of fusion rules, the existence of such a factorization is obvious since, by the Chinese remainder theorem, finite Abelian groups admit such a factorization. But to establish the factorization at the level of TQFTs, it is also necessary to show that anyons in the different factors braid trivially. To show this, first note that for any anyon $a_i$ in ${\cal T}_{p_i}$, the order $m_i$ of $a_i$ must divide the total number of anyons in ${\cal T}_{p_i}$ by Lagrange's theorem. This means that $m_i$ is a power of $p_i$, and thus $\mathrm{gcd}(m_i, m_j) = 1$ for $i \neq j$. Given anyons $a_1$ and $a_2$ in ${\cal T}_{p_1}$ and ${\cal T}_{p_2}$ with respective orders $m_1$ and $m_2$, the braiding must satisfy 
\bea
B(a_1,a_2)^{m_1} = B(a_1,a_2)^{m_2} = 1
\eea
by the multiplicative property of the braiding phase. But since $m_1$ and $m_2$ are coprime, the only solution to this is the trivial phase. This justifies the decomposition. 

To prove the factorization of the Frobenius-Schur exponent, we note that the spins satisfy $\theta(a^{m}) = \theta(a)^{m^2}$. Considering an anyon $a_i$ of order $m_i$ in ${\cal T}_{p_i}$, we conclude that $\theta(a_i)^{m_i^2} = 1$ and thus, since $m_i$ is a power of $p_i$, $N_i$ is as well. This tells us that $\mathrm{gcd}(N_i , N_j ) = 1$ for $i \neq j$, from which $N_\mathrm{FS}=N_1 N_2 \dots N_k$ follows.\hfill $\square$
\newline

Having shown that any Abelian TQFT $\cal T$ admits a factorization as in (\ref{eq:AbTFTfact}), we now show the following,

{\lem $\cal T$ admits a gapped boundary if and only if all the factors ${\cal T}_{p_i}$ do. \label{lem:gappedfact}}
\newline\newline
Denote the total number of anyons in $\cal T$ by $|{\cal T}|$. Recall that $\cal T$ admits a gapped boundary if and only if there exists a Lagrangian subgroup $L$, i.e., a subgroup of $|L|=\sqrt{|{\cal T}|}$ lines with trivial spins. This in particular requires that $|{\cal T}|$ is a perfect square, in which case the orders of the  theories ${\cal T}_{p_i}$ appearing in the prime factor decomposition are also perfect squares. 
Clearly such an $L$, if  exists, admits a decomposition as $L= L_{p_1} \times \dots \times L_{p_k}$ by the exact same reasoning as for $\cal T$.  Furthermore, we see that $|L_{p_i}| = \sqrt{|{\cal T}_{p_i}|}$ and, since the lines in $L$ have trivial spin, those in $L_{p_i}$ do as well. Thus each $L_{p_i}$ serves as a Lagrangian subgroup for ${\cal T}_{p_i}$, proving the forward direction of the theorem. Conversely, given a Lagrangian subgroup for each ${\cal T}_{p_i}$, it is easy to see that the product of these subgroups gives a Lagrangian subgroup for $\cal T$, since lines in different factors have trivial braiding.\hfill $\square$

We are now in a position to prove Theorem \ref{thm:compobs}. We begin by proving the forward direction, namely that the existence of a gapped boundary for $\cal T$ implies $\xi_n({\cal T})= 1$ for all $n$ such that ${\rm gcd}\left(n , {N_\mathrm{FS} \over {\rm gcd}(n,N_\mathrm{FS})} \right) =1$. By Lemma \ref{lem:factlem}, the theory $\cal T$ admits the  factorization (\ref{eq:AbTFTfact}) with $N_\mathrm{FS}= N_1\dots N_k$. 
The condition on $n$ is equivalent to requiring that for any $i$, either $\mathrm{gcd}(n, N_i) = 1$ or $N_i \, | \, n$. By Lemma \ref{lem:gappedfact} the existence of a gapped boundary for $\cal T$ means that all factors ${\cal T}_{p_i}$ have gapped boundaries as well. Now consider any factor ${\cal T}_{p_i}$; as we have reviewed in the previous subsection, for $\mathrm{gcd}(n,N_i)=1$, the existence of a gapped boundary implies $\xi_n({\cal T}_{p_i}) = 1$ \cite{Ng:2018ddj}. On the other hand, if $N_i \, | \, n$, then $\xi_n({\cal T}_{p_i}) =1$ trivially. Either way, when a gapped boundary exists we see that the relevant higher central charges for the prime factors are trivial. Noting that 
\bea
\xi_n({\cal T}) = \xi_n({\cal T}_{p_1})  \xi_n({\cal T}_{p_2}) \dots  \xi_n({\cal T}_{p_k}) 
\eea
then completes the proof of the forward direction. 

For the converse direction, assume that $\xi_n({\cal T})= 1$ for all $n$ such that ${\rm gcd}\left(n , {N_\mathrm{FS} \over {\rm gcd}(n,N_\mathrm{FS})} \right) =1$. In particular, we can consider $n$ of the form $n= N_1\dots N_{r-1} N_{r+1} \dots N_k \tilde n$ with $\mathrm{gcd}(\tilde n, N_r)=1$. Since all $N_i$ except for $N_r$ divide $n$, we have 
\bea
\xi_n({\cal T}) = \xi_n({\cal T}_{p_r}) ~.
\eea
If we now take $\tilde n$ to scan over all totatives of $N_r$, then $n$ also scans over all totatives of $N_r$. Thus if $\xi_n({\cal T}) = 1$ for all such $n$, we conclude that $\xi_n({\cal T}_{p_r})=1$ for all $\mathrm{gcd}(n, N_r) = 1$. For theories where the number of anyons is a prime power, as is the case for ${\cal T}_{p_r}$, it is already known that this implies the presence of a gapped boundary (see \cite[Appendix A.7]{drinfeld2010braided} for a proof). Repeating this for all $r$ and utilizing Lemma \ref{lem:gappedfact} then completes the proof.

\subsubsection*{Summary on Topological Boundary Conditions of Abelian TQFTs}

We summarize the discussions up to this point by the following statement. 
If we work modulo invertible field theories (such as the $(E_8)_1$ CS theory), then the following conditions for a bosonic, Abelian TQFT are equivalent:
\begin{itemize}
\item It admits a topological boundary condition.
\item It has a Lagrangian subgroup, i.e., a non-anomalous subgroup $L$ of the one-form symmetry $G$ satisfying $|L|^2 =|G|$.
\item It is an Abelian Dijkgraaf-Witten gauge theory.
\item  $\xi_n = { \sum_a \theta(a)^n \over| \sum_a \theta(a)^n|}=1$ for all $n$ such that ${\rm gcd}\left( n, {2 |G| \over {\rm gcd}(n,2|G|)}\right) =1$.
\end{itemize} 

\subsection{Another Point of View}\label{sec:anotherpov}

In Section \ref{sec:main} we established that if $\cal T$ has a gapped boundary, then $Z_{\cal T}[M]$ is positive on every closed oriented three-manifold $M$ with $\text{gcd}( |H_1(M)|  , |G|)=1$. (More precisely, this is true in some particular scheme which always exists if $c_-=0$. This is the scheme where the partition functions are topological invariants and framing-independent.)

There is another point of view on this result which partly generalizes to non-Abelian theories. The idea is to think of $Z_{\cal T}[M]$ as the $S^3$ partition function of some different (Abelian) theory, with a possibly different number of anyons. Assuming the existence of a Lagrangian subgroup in the original theory, one can prove the existence of a Lagrangian subgroup in the auxiliary theory, which guarantees that it has a vanishing chiral central charge mod 8 and hence that its $S^3$ partition function is positive. Translating back to the original theory proves the positivity of $Z_{\cal T}[M]$. This point of view, besides partly generalizing to non-Abelian theories, also allows to establish various properties of the partition functions  $Z_{\cal T}[M]$. For instance, the phase of $Z_{\cal T}[M]$ is always an 8th root of unity (whether or not $c_-=0$ in the original theory).

\subsubsection*{$K$-matrix}
To see how this works, first recall   that any Abelian TQFT can be presented as a Chern-Simons theory with some
$K$-matrix $\bfK$ \cite{Belov:2005ze} (see also \cite{WALL1963281,https://doi.org/10.1112/blms/4.2.156,Nikulin_1980,Stirling:2008bq}). These theories have Lagrangians
\ie
 {\cal L} = {1\over 4\pi} K_{IJ} \, a_I\wedge \mathrm{d} a_J~, 
 \fe
where the $a_I$ are $U(1)$ gauge fields and $I=1,...,|\bfK|$.
The matrix $\bfK$ is symmetric and integral. For the theory to be bosonic,  we furthermore require that the diagonal entries of $\bfK$ are even. The anyons in these theories are labeled by $|\bfK|$-dimensional integer-valued vectors ${\bf m} \in \bZ^{|\bfK|}$, and fusion of two anyons ${\bf m}$ and ${\bf n}$ corresponds to addition of vectors
${\bf m} + {\bf n}$. The associated topological spins are given by
\ie
\theta({\bf m}) = e^{\pi i \, {\bf m}^\mathrm{T} {\bfK}^{-1} {\bf m}}~. \label{spin.k.matrix}
\fe
Of course, not all integer-valued vectors ${\bf m}$ describe independent anyons. There are only finitely many anyons which are independent and furnish a non-degenerate braiding matrix. 
Indeed, the braiding phase is calculated from the topological spins as usual for Abelian theories,
\ie
B({\bf m},{\bf n}) =e^{2\pi i \, {\bf m}^\mathrm{T}{\bf K}^{-1} {\bf n} }~,
\fe
so if we shift ${\bf m} \to {\bf m} + \bfK \cdot {\bf \tilde m}$ for arbitrary ${\bf\tilde m}\in \mathbb{Z}^{|\bfK|}$, the braiding is left unchanged. The topological spin is also invariant under such a shift,\footnote{This is why $\bfK$ has to have even integer entries on the diagonal. Otherwise, there would be transparent spin 1/2 anyons.  Such theories with a transparent spin 1/2 anyon are called spin TQFTs. } and therefore the space of anyon labels, or more precisely the Abelian group of anyons, is
\ie
G = {\mathbb{Z}}^{|\bfK|}/{\bf K}\cdot {\mathbb{Z}}^{|\bfK|}~.
\fe
In particular, there are $|G|=|\det {\bf K}|$ independent anyons.

Let us evaluate the $L(1,1) \simeq S^3$ partition function in such a $K$-matrix theory.
To do so, we need to perform the sum~\eqref{LensP} 
\ie
Z\big[L(1,1)\big] = {1\over |\det {\bf K}|}\sum_{{\bf m}\in {\mathbb{Z}}^{|\bfK|}/{\bf K}\cdot {\mathbb{Z}}^{|\bfK|}}  e^{\pi i\, {\bf m}^\mathrm{T} {\bf K}^{-1} {\bf m}}~.
\fe
By referring to~\cite[Theorem 1]{styer1984evaluating} (see also~\cite{Belov:2005ze}), one finds
\begin{equation}\label{threes}
	{1\over |\det {\bf K}|}\sum_{{\bf m}\in {\mathbb{Z}}^{|K|}/{\bf K}\cdot {\mathbb{Z}}^{|K|}}  e^{\pi i \, {\bf m}^\mathrm{T} {\bf K}^{-1} {\bf m}}= \frac{1}{\sqrt{|\det \bfK|}} e^{{2\pi i\over 8} \mathrm{sgn}({\bf K}) }~,
\end{equation}
where $\mathrm{sgn}({\bf K})$ is the signature of the symmetric matrix ${\bf K}$. This is the familiar statement that the chiral central charge is the signature of the matrix $\bfK$. 

We would now like to consider the partition function of the theory on a general 3-manifold obtained by surgery on a link with linking matrix $\bfL$ (a brief review of surgery is given in Appendix \ref{app:RT}). Like $\bfK$, the linking matrix $\bfL$ is an integral-valued symmetric matrix, though now the diagonal does not have to consist of even numbers. For Abelian theories the partition function depends only on the linking matrix, since for any two simple anyons the fusion channel is unique. The resulting manifold has $H_1(M)$ isomorphic to 
${\mathbb{Z}}^{|\bfL|}/{\bfL}\cdot {\mathbb{Z}}^{|\bfL|}$, and in particular $|H_1(M)|=|\det \bfL|$.

Using the methods reviewed in Appendix \ref{app:RT}, one finds that the partition function of the theory with $K$-matrix $\bfK$ on a 3-manifold with linking matrix $\bfL$ takes the form (\textit{cf.} equation \eqref{RT})
\begin{equation}\label{partfun}
	Z_{\bfK}[\bfL]={1\over |\det \bfK|^{|\bfL|/2+1/2}}\sum_{{\bf m}\in {\mathbb{Z}}^{|\bfK||\bfL|}/ ({\bfK}\otimes \mathbbm{1}) \cdot {\mathbb{Z}}^{|\bfK||\bfL|} }e^{\pi i \, {\bf m}^\mathrm{T} ({\bfK}^{-1} \otimes {\bfL}) {\bf m}}~.
\end{equation}
It is difficult to evaluate this in closed form. However, we now show that under certain conditions on the matrices $\bfK$ and $\bfL$ we can reinterpret \eqref{partfun} as the $S^3$ partition function of an auxiliary theory, which can then be computed using (\ref{threes}).  
Similar calculations have been done in \cite{Guadagnini:2014mja}.

\subsubsection*{The Auxiliary Theory}
The most general situation in which we can recast \eqref{partfun} as the $S^3$ partition function of an auxiliary theory is when the matrices $\bfK\otimes \mathbbm{1}$ and $\mathbbm{1}\otimes \bfL$ are \textit{coprime}. Two $n\times n$ integral matrices $C$ and $D$ (both assumed to have non-zero determinant) are called coprime if there exist integral matrices $A$ and $B$ such that~\cite{maass1954lectures}
\begin{equation}
	\label{Bezout} AD^\mathrm{T} - BC^\mathrm{T} =\mathbbm{1}~.
\end{equation}
When $C D^\mathrm{T}$ is an even symmetric matrix (this is often referred to as a symmetric pair), there exists an integral symplectic matrix whose lower row is $(C,D)$. In other words, there exists a ``preferred'' choice of $A$ and $B$ in~\eqref{Bezout} such that
\begin{equation}
	A^\mathrm{T}D-C^\mathrm{T}B=\mathbbm{1}~,
	\quad A^\mathrm{T}C=C^\mathrm{T}A~,\quad B^\mathrm{T}D=D^\mathrm{T}B~. \label{symplectic}
\end{equation}

In the context we are interested in, we take $C=\bfK\otimes \mathbbm{1}$ and $D=\mathbbm{1}\otimes \bfL$. 
These matrices commute and each of them is symmetric, and hence they form a symmetric pair. This symmetric pair is even because the elements on the diagonal of $\bfK$ are even. Hence coprimality is equivalent to demanding the existence of  integral matrices $A$ and $B$ satisfying (\ref{symplectic}). 
In Appendix \ref{app.Abelian}, we prove that such $A$ and $B$ exist if and only if $\mathrm{gcd}(|\det \bfK|,|\det \bfL|)=1$. Since in the context of surgery $|\det \bfL|$ is the order of $H_1$, the condition $\mathrm{gcd}(|\det \bfK|,|\det \bfL|)=1$ is precisely the one we are interested in for the question of gapped boundaries, namely $\text{gcd}( |G|, |H_1(M)| )=1$.

Assuming comprimality, we now interpret \eqref{partfun} as the $S^3$ partition function of an auxiliary Abelian theory. In particular, the auxiliary theory is one whose anyons generate the Abelian group ${\mathbb{Z}}^{|\bfK||\bfL|}/({\bf K}\otimes \mathbbm{1})\cdot {\mathbb{Z}}^{|\bfK||\bfL|}$, with spins 
\begin{equation}
	\theta({\bf m}) = e^{\pi i\, {\bf m}^\mathrm{T} ({\bfK}^{-1}\otimes{\bfL}) {\bf m}}~. \label{theta}
\end{equation} For this data to define a legitimate theory, we must check that the braiding is a bilinear map with $\theta$ being its quadratic refinement, and that the resulting braiding matrix is non-degenerate. Regarding the first check, it is immediate that $\theta(zxy)={\theta(zy)\theta(xy)\theta(zx)\over \theta(x)\theta(y)\theta(z)}$, and hence the braiding is bilinear. As for non-degeneracy, we have to check that~\eqref{Ortho} holds. This requires that
\begin{equation}
	\sum_{{\bf c}\in {\mathbb{Z}}^{|\bfK||\bfL|}/ ({\bfK}\otimes \mathbbm{1}) \cdot {\mathbb{Z}}^{|\bfK||\bfL|} }  e^{ 2 \pi i \, {\bf a}^\mathrm{T} ({\bfK}^{-1}\otimes{\bfL}) {\bf c} \, +\,  2 \pi i \, {\bf b}^\mathrm{T} ({\bfK}^{-1}\otimes{\bfL}) {\bf c} } = |\det \bfK|^{|\bfL|} \,\delta\left({\bf a} + {\bf b}\right)~,
\end{equation}
where
\ie
\delta\left({\bf a}\right) = \begin{cases}
	1 & \text{if } {\bf a} \in ({\bfK}\otimes \mathbbm{1}) \cdot {\mathbb{Z}}^{|\bfK||\bfL|}  \\
	0 & \text{otherwise}
\end{cases}~.
\fe
To show this, we can trivially rewrite the LHS as 
\ie
\sum_{{\bf c}\in {\mathbb{Z}}^{|\bfK||\bfL|}/ ({\bfK}\otimes \mathbbm{1}) \cdot {\mathbb{Z}}^{|\bfK||\bfL|} }  e^{2 \pi i \, [\mathbbm{1}\otimes{\bfL} \cdot ({\bf a} + {\bf b})]^\mathrm{T} ({\bfK}^{-1}\otimes\mathbbm{1}) {\bf c}} = |\det \bfK|^{|\bfL|} \, \delta\left( (\mathbbm{1}\otimes \bfL) \cdot ({\bf a}+ {\bf b})\right)~,
\fe
where the sum was evaluated by interpreting it in terms of the braiding matrix of a theory consisting of $|\bfL|$ copies of the original theory defined by $\bfK$. Since the original theory was assumed to have a non-degenerate braiding, we were able to evaluate the sum using \eqref{Ortho}.

If we could now show that $\delta\left( (\mathbbm{1}\otimes \bfL) \cdot {\bf a}\right) = \delta\left({\bf a}\right)$, we would have successfully proven the non-degeneracy of the braiding for the auxiliary theory. To do so, suppose that 
$(\mathbbm{1}\otimes \bfL) \cdot {\bf a} = (\bfK\otimes \mathbbm{1}) \cdot {\bf m}$ for some integer vector $\bf m$. Setting $C=\bfK\otimes \mathbbm{1}$ and $D=\mathbbm{1}\otimes \bfL$, by the assumption of coprimality there exist integer matrices $A$ and $B$ such that
\ie
 A^\mathrm{T} D - C B=\mathbbm{1}~,
\fe
where we have used the fact that $C$ is symmetric. Acting with both sides of the equation on the vector $\bf a$, we find
$ A^\mathrm{T} C \cdot {\bf m} - C B \cdot {\bf a}= {\bf a}$.
But we also know that $ A^\mathrm{T} C=C A$ (from (\ref{symplectic}) and the fact that $C$ is symmetric) and hence
\ie
 (\bfK\otimes \mathbbm{1}) \left( A \cdot {\bf m} - B \cdot {\bf a} \right) = {\bf a}~,
 \fe
which means that $\delta\left( (\mathbbm{1}\otimes \bfL) \cdot {\bf a}\right) = \delta\left({\bf a}\right)$ as we wanted.

We have thus learned that the spins given in (\ref{theta}) give a consistent quadratic refinement of a bilinear map on ${\mathbb{Z}}^{|\bfK||\bfL|}/({\bf K}\otimes \mathbbm{1})\cdot {\mathbb{Z}}^{|\bfK||\bfL|}$. It follows that the auxiliary theory proposed is a legitimate Abelian TQFT, and that \eqref{partfun} can be interpreted as its $S^3$ partition function. 
We may then use the general $S^3$ result \eqref{threes} to conclude that 
\ie
\sum_{{\bf m} \in {\mathbb{Z}}^{|\bfK||\bfL|}/ ({\bfK}\otimes \mathbbm{1}) \cdot {\mathbb{Z}}^{|\bfK||\bfL|} }e^{\pi i \, {\bf m}^\mathrm{T} {\bfK}^{-1}\otimes{\bfL} {\bf m}}=\sqrt{|\det \bfK|^{|\bfL|}}\ \xi_{\bfK, \bfL}~,
\fe
for some 8th root of unity $\xi_{\bfK,\bfL}$. 
It follows that
\begin{equation}
\label{eq:genZKL}
	Z_{\bfK}[\bfL]={1\over |\det \bfK|^{|\bfL|/2+1/2}}\sqrt{|\det \bfK|^{|\bfL|}}\ \xi_{\bfK, \bfL}=
	{1\over |\det \bfK|^{1/2}}\ \xi_{\bfK, \bfL}~.
\end{equation}
This proves that  the phase of $Z_{\bfK}[\bfL]$ is always an 8th root of unity, consistent with the fact that 8 copies of any Abelian theory has a gapped boundary \cite[Section 5.3]{davydov2013witt}. We have also determined the absolute value of these partition functions. 

We now prove that if the original theory defined by $\bfK$ has a gapped boundary, then $\xi_{\bfK, \bfL}=1$. To do so, we assume that the original theory has a Lagrangian subgroup of anyons ${\cal A} \subset \mathbb{Z}^{|\bfK|}/\bfK \cdot \mathbb{Z}^{|\bfK|}$, where $\theta(a)=1$ for all $a \in{\cal A}$ (and in particular this means that $B(a,b)=1$ for all $a,b\in {\cal A}$). Then in the auxiliary theory, ${\cal A}^{\otimes |\bfL|}$ is also a Lagrangian subgroup.

To prove ${\cal A}^{\otimes |\bfL|}$ is Lagrangian, we have to show that it has the correct size and moreover the spin of all anyons in this subgroup is trivial. The first property follows easily since $|{\cal A}^{\otimes |\bfL|}| = |\cA|^{|\bfL|} = \sqrt{|\det \bfK|^{|\bfL|}}$. For the second property, note that an arbitrary anyon in this subgroup can be represented by a vector ${\bf m} = {\bf m}_1 \oplus \cdots \oplus {\bf m}_{|\bfL|} \in \bZ^{|\bfK||\bfL|}$ for some ${\bf m}_i$ representing an anyon in $\cA$. To show that the spin of ${\bf m}$ given by \eqref{theta} is trivial, note that
\begin{equation}
	\frac12 {\bf m}^\mathrm{T} (\bfK^{-1} \otimes \bfL) {\bf m} = \sum_{ i < j} L_{ij} \, {\bf m}_i^\mathrm{T} \bfK^{-1} {\bf m}_j + \frac12 \sum_{ i} L_{ii} \, {\bf m}_i^\mathrm{T} \bfK^{-1} {\bf m}_i \in \bZ \,.
\end{equation}
Above we have used $\frac12 {\bf m}_i^\mathrm{T} \bfK^{-1} {\bf m}_i \in \bZ$, which is true because $\cA$ is Lagrangian.
Thus the auxiliary theory must have vanishing chiral central charge, and hence its $S^3$ partition function, namely (\ref{eq:genZKL}), must have $\xi_{\bfK, \bfL}=1$. This gives an alternative proof of Theorem \ref{theorem}.

\section{Non-Abelian TQFTs}\label{sec:nonAbel}

In this section we discuss the existence of topological boundary conditions for non-Abelian unitary 2+1d TQFTs. As before we set $c_-=0$ (which means that the discussion applies to theories with $c_-\in 8\mathbb{Z}$ after an appropriate stacking with copies of the invertible $(E_8)_1$ CS theory) and work with framing-independent topological  theories.

\subsection{Lagrangian Algebra Anyons} \label{sec:Lagalg}
For general non-Abelian TQFTs, the question of whether a gapped boundary exists cannot be reduced to a question about a Lagrangian subgroup of anyons. However, there is a related object known as a \textit{Lagrangian algebra anyon} which will allow us to make statements about gapped boundaries in the non-Abelian case. 

Let us assume that a gapped boundary exists in a given non-Abelian TQFT. Then we can cut out a small cylindrical tube and introduce the topological boundary condition on the surface of the tube. Since the boundary condition is topological we can change the radius of the cylindrical tube at will. Shrinking it must therefore define a line defect, which is equivalent to a direct sum of simple anyons (see Figure~\ref{fig:tube})
\begin{figure}
	\centering
	\begin{tikzpicture}[scale = .9]
		\node[cylinder, draw, shape aspect=.5, 
		cylinder uses custom fill, cylinder end fill=gray!30, 
		minimum height=.8cm,
		cylinder body fill=red!25, opacity=0.5, 
		scale=4, rotate=90]{};
		\draw[ -> ,ultra thick] (1.5,0) --node[midway,anchor = south]{shrink} ++(2.5,0);
		\draw[ultra thick, red!70] (4.7,-1.4) -- (4.7,1.4);
		\node at (5,1.5) {$\cA$};
	\end{tikzpicture}
	\caption{Anyon $\cA$ from gapped boundary condition.}
	\label{fig:tube}
\end{figure}
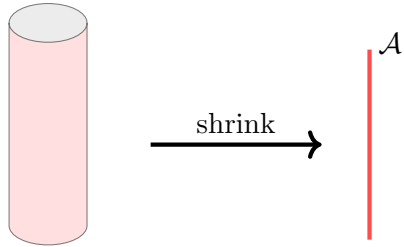
\begin{equation}
\label{eq:Lagalgdef}
	{\cA} = \bigoplus_{a\,\in\, \mathcal{I}} Z_{0a}\, a~,
\end{equation}
where $Z_{0a}$ are some non-negative integers and $\mathcal{I}$ is the set of simple anyons.

The vector $Z_{0a}$ obeys many nice properties. 
Below we will show that $Z_{0a}$ is an eigenvector of the $S$- and $T$-matrices of the TQFT with eigenvalue 1.  
A corollary is  that the symmetric matrix $Z_{ab}:=Z_{0a}Z_{0b}$  commutes with the $S$ and $T$ matrices.

\begin{figure}[!tbp]
\begin{center}

\includegraphics{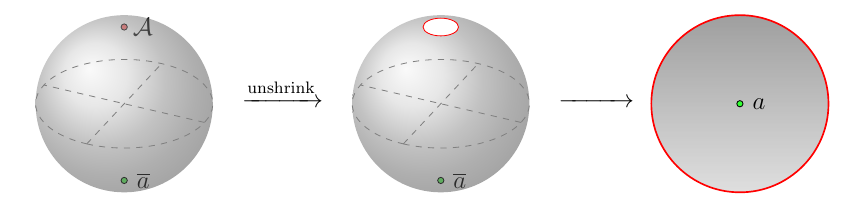}

\caption{The anyons $\overline{a}$ and $\cA$ wrap an $S^1$ (not shown) and are located at points on $S^2$. Replacing the algebra anyon $\cA$ with an empty tube turns the $S^2$ into a disk, and the partition function of this configuration gives the dimension of the Hilbert space on $D^2$ with the insertion of $a$.}
\label{fig:Z0adimH}
\end{center}
\end{figure}

We can compute the integers $Z_{0a}$ by considering the $S^2\times S^1$ partition function with the insertion of parallel anyons $\cA$ and $\bar{a}$ along $S^1$~\cite{Witten:1988hf}, as shown in Figure \ref{fig:Z0adimH}.
On the one hand, this partition function is equal to the dimension of the Hilbert space on $S^2$ punctured by $\cA$ and $\bar{a}$, which is equal to $\mathrm{dim}\, \mathrm{Hom}(\cA \otimes \bar{a},1) = \mathrm{dim}\, \mathrm{Hom}(\cA,a) =Z_{0a}$. On the other hand, viewing $\cA$ as the empty cylindrical tube, this configuration is topologically equivalent to the solid torus $D^2\times S^1$ with the gapped boundary condition on its boundary and the insertion of the anyon $a$ along $S^1$. From this point of view the partition function is equal to the dimension of the disk Hilbert space punctured by $a$, and equating the two results gives
\begin{equation}
\label{eq:Z0adimH}
	Z_{0a} = \mathrm{dim}\, \mathcal{H}(D^2;a)~.
\end{equation}
By the state/operator correspondence, the Hilbert space $\mathcal{H}(D^2;a)$ is the same as the space of operators living at the intersection of the line $a$ with the boundary. Therefore $Z_{0a}$ counts the number of distinct ways that $a$ can end on the gapped boundary. In particular $Z_{0a} \neq 0$, if and only if the line $a$ can end (condense) on the boundary.

The anyon $\cA$ defined above has various special properties that we will explore in the remainder of this subsection.
Any anyon with these properties is known as a \textit{Lagrangian algebra} \cite{davydov2013witt}, so we see that the existence of a gapped boundary implies the existence of a Lagrangian algebra anyon. By introducing the notion of gauging the algebra anyon, we will see how the original TQFT turns into a trivial theory.  

\subsubsection*{$Z_{00}=1$}
First we show that the trivial anyon must be contained in $\cal A$. To prove this, we must show that there exists a non-zero morphism between $\cA$ and the trivial anyon, meaning that the Lagrangian algebra anyon $\cA$ can end. This is most easily seen by replacing $\cA$ with an empty tube, and noting that we can cap off the empty tube. For instance, we can imagine putting the gapped boundary condition on the boundary of a three-dimensional ball which is topologically equivalent to an elongated cigar. By shrinking the width of the cigar, the configuration can be interpreted as the anyon $\cA$ morphing into the identity line on both ends. This means that $\mathrm{Hom}({\cA},1)$ is non-empty, and thus $Z_{00} \neq 0$.

One can further argue that $Z_{00}=\mathrm{dim}\, \mathcal{H}(D^2)=1$ for simple gapped boundary condition. Indeed, we will now show that if the disk Hilbert space is not one-dimensional, then the boundary condition is not simple and can be decomposed into the direct sum of $Z_{00}$ simple gapped boundary conditions.
To show this, first recall that by the state/operator correspondence $\mathcal{H}(D^2)$ can be identified with the algebra of boundary (point) operators. These boundary operators are topological and live on a two-dimensional surface. Hence their ``OPE" defines a commutative Frobenius algebra\footnote{For instance, the unit and the trace of this algebra is given by the 3-ball partition function with boundary insertions of these local operators.} analogous to the algebra of point operators in 1+1d TQFTs (see for instance~\cite{Moore:2006dw}). In other words, $\mathcal{H}(D^2)$ can be identified with the $S^1$ Hilbert space of a 1+1d TQFT constructed from the topological boundary condition. When we fix the three-dimensional bulk to be a genus-$g$ handlebody, we can take the topological boundary condition to be this 1+1d TQFT on a genus-$g$ surface.

Assuming unitarity (reflection-positivity), the Frobenius algebra $\mathcal{H}(D^2)$ must be semisimple (separable)~\cite{Durhuus:1993cq,Moore:2006dw}. Therefore it must contain a complete set of projection operators (idempotents)
\begin{equation}
	\varepsilon_1, \varepsilon_2, \dots, \varepsilon_{Z_{00}} \in  \mathcal{H}(D^2)~, \qquad \varepsilon_i \varepsilon_j = \delta_{ij} \varepsilon_i~.
\end{equation}
When inserted on the boundary, the topological boundary operator $\varepsilon_i \in \mathcal{H}(D^2)$ will project the gapped boundary condition onto its $i$-th simple component.
In other words, smearing (or, condensing) $\varepsilon_i$ over the boundary, which is equal to inserting one, defines a boundary condition that is contained in the original boundary condition as a summand.
Hence we have established that in a unitary theory any topological boundary condition decomposes into simple boundary conditions with one-dimensional disk Hilbert spaces. 

\subsubsection*{$TZ=Z$}
Next we show that $Z_{0a}$ must satisfy $(TZ)_{0a}=Z_{0a}$. (This also means that $Z_{ab}:=Z_{0a}Z_{0b}$ commutes with the $T$-matrix.) 
Since the boundary is gapped, Dehn twisting leaves the partition function invariant. Hence for all $a$ with $Z_{0a}\neq 0$ we have $\theta(a)=1$. In other words, the anyons that make up ${\cal A}$ must have zero spin, and hence
\begin{equation}
	\sum_b T_{ab} Z_{0b} = Z_{0a}~. \label{TZ}
\end{equation}

\subsubsection*{$SZ=Z$}

\begin{figure}[!tbp]
	\begin{center}
		\includegraphics{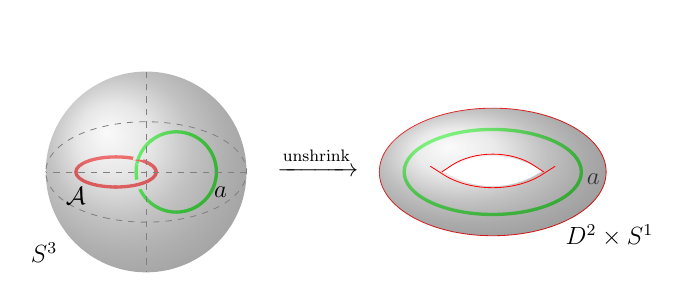}
		\caption{The Hopf link between $a$ and $\cA$ can be evaluated to $(SZ)_{0a}$. On the other hand, replacing $\cA$ by an empty tube gives a solid torus with $a$ wrapping the non-contractible cycle, which can be evaluated to the dimension of the Hilbert space on $D^2$ with an insertion of $a$, i.e. $Z_{0a}$.}
		\label{fig:SZZ}
	\end{center}
\end{figure}

Furthermore, $Z_{0a}$ must satisfy $(SZ)_{0a}=Z_{0a}$. (As we remarked earlier, from this also follows that  $Z_{ab}:=Z_{0a}Z_{0b}$ commutes with the $S$ matrix.) To see this, consider the $S^3$ partition function with the insertion of the Hopf link between $\cA$ and $a$, as shown in Figure \ref{fig:SZZ}. This is equal to the  Hopf link amplitude, times the $S^3$ partition function without any insertions (the latter of which is equal to $S_{00}$  in theories with a vanishing chiral central charge mod 8)~\cite{Witten:1988hf}. Splitting the algebra anyon $\cA$ via (\ref{eq:Lagalgdef}) and then applying (\ref{Smat}), we get
\begin{equation}
\label{eq:SZ}
	Z\big[  S^3;\mathrm{Hopf}(\cA,a) \big] = \sum_b S_{ab} Z_{0b}~.
\end{equation}
On the other hand, thinking of $\cA$ as the empty tube this configuration is topologically equivalent to the solid torus with the gapped boundary condition on its boundary and the anyon $a$ inserted along its longitude. In this case the partition function evaluates to the dimension of the disk Hilbert space punctured by $a$, and using (\ref{eq:Z0adimH}) we conclude that
\begin{equation}
	\sum_b S_{ab} Z_{0b} = Z_{0a}~. \label{SZ}
\end{equation}
We have therefore proven that  the matrices $S$ and $T$ have a common non-negative integer eigenvector $Z_{0a}$ with eigenvalue $1$.  Furthermore, $Z_{0a}$ can be shown to obey  \cite{Lan:2014uaa}:
\ie
Z_{0a} Z_{0b} \leq \sum_c N_{ab}^c Z_{0c}\,.\label{ZZNZ}
\fe

The conditions  \eqref{TZ}, \eqref{SZ}, and \eqref{ZZNZ} are only necessary conditions for the Lagrangian algebra anyon, but not sufficient \cite{Kawahigashi:2015lxa}. 
More generally, for the genus-$g$ surface $\Sigma_g$, there exist a vector $\ket{\cA} \in \mathcal{H}(\Sigma_g)$ that is preserved by all genus-$g$ mapping class group (MCG) transformations. The state $\ket{\cA}$ is defined by the path integral on $\Sigma_g \times [0,1]$, where we put the topological boundary condition on $\Sigma_g \times \{0\}$. The path integral prepares the \emph{boundary state} $\ket{\cA}$ on the Hilbert space on the other side of the interval $\Sigma_g \times \{1\}$.
Since MCG transformations do not change the topology of the boundary, they act trivially on the topological boundary condition. This shows that indeed $\ket{\cA}$ is a singlet of the MCG representation. For $g=1$ we have $\ket{\cA} = \sum_a Z_{0a} \ket{a}$, where $\ket{a}$ is the state prepared by the path integral on the solid torus with an insertion of anyon $a$ wrapping its non-contractible cycle.

\subsubsection*{Quantum dimension $\mathrm{dim}(\cA) $}
The existence of the eigenvector $Z_{0a}$ is very constraining. In particular, such an eigenvector can only exist if $c_-=0 \mod{8}$. To see this, note  that $S$ and $T$ satisfy
\begin{equation}
	(ST)^3 = e^{2\pi i \frac{c_-}{8}} S^2~.
\end{equation}
Then we can act on both sides of this equation on $Z$, and find that this is only consistent if $c_-=0 \mod{8}$.

Moreover, setting $a=0$ in \eqref{SZ}, we find that the quantum dimension of $\cA$, denoted by $\mathrm{dim}(\cA) = \sum_aZ_{0a}d_a$, is equal to the total quantum dimension
\begin{equation}
	\mathrm{dim}(\cA) = \sqrt {\sum_a d_a^2}~.\label{Lag.prop}
\end{equation}
Here we have made use of (\ref{eq:defofS}) (fixing $b$ to the identity tells us that $d_a = {S_{0a} S_{00}^{-1}}$), together with (\ref{eq:totD}). 
These statements can be translated to facts about the $S^3$ partition function. Indeed, the 3-sphere can be obtained by gluing two solid tori along their boundary with an $S$-transformation, and thus
\begin{equation}\label{qdimrel}
	Z[S^3] = S_{00} = \frac{1}{\mathrm{dim}(\cA)}~.
\end{equation}

Another way to obtain~\eqref{qdimrel} is to consider an unknot of $\cA$ inside $S^3$. On the one hand the result is by definition $Z[S^3] \mathrm{dim}(\cA)$. On the other hand, we can blow up the unknot and obtain a disc with topological boundary condition and no insertion. The partition in this presentation is manifestly 1 (note that $Z_{00}=1$). We have therefore derived \eqref{qdimrel}.

Note that all of these constraints nicely generalize facts that we have explained in great detail about the Abelian theories. For instance, in the Abelian case $\cal A$ is nothing more than a direct sum of the lines in the Lagrangian subgroup, each appearing with multiplicity 1. Above we have rederived the facts that the spins of the lines in the Lagrangian subgroup all vanish and that the dimension of the Lagrangian subgroup must be the square root of the total number of anyons.

\subsubsection*{$F^{{\cal A}{\cal A}{\cal A}}_{{\cal A}}$ and $R^{{\cal A}{\cal A}}_{{\cal A}}$}

\begin{figure}[!tbp]
	\begin{center}
		\includegraphics{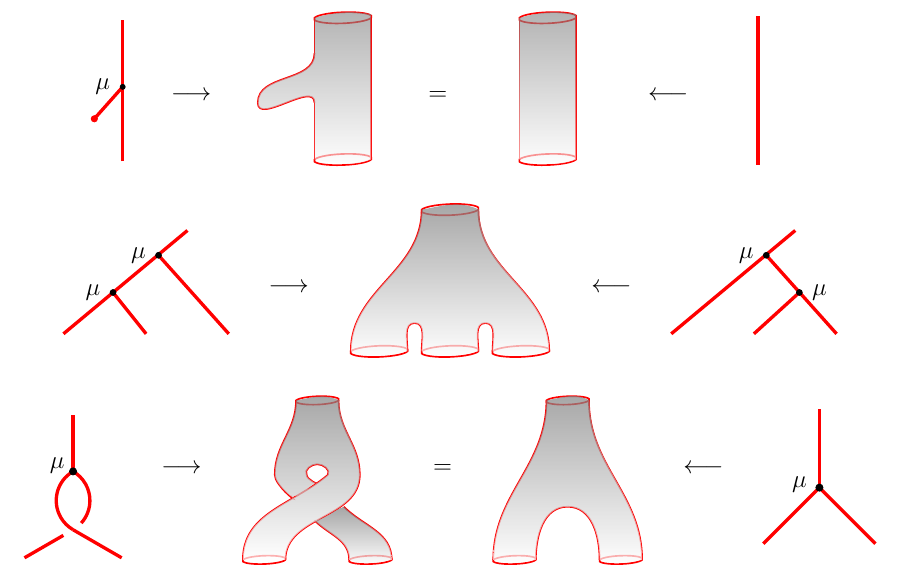}
		\caption{Replacing the $\cA$ anyons with empty tubes makes it clear that fusion and braiding are trivial. This gives a geometric interpretation of the defining axioms of a Lagrangian algebra $\cal A$.}
		\label{fig:braidingfusionA}
	\end{center}
\end{figure}

Additional constraints on the Lagrangian algebra $\cal A$ arise if we consider carving out junctions of cylinders (pairs of pants), such that the boundary is our gapped boundary condition. Clearly since the boundary is topological, we can do arbitrary smooth transformations of these junctions and find ``trivial" $F$ and $R$ moves for $\cA$, as shown in Figure \ref{fig:braidingfusionA}. More precisely, let
$\ket{\mu} \in V_\cA^{\cA\cA}$
be the junction vector/operator between three ${\cA}$ specified by putting the gapped boundary condition on a pair of pants. Then we have (\textit{cf.} equation \eqref{f-matrices})
\begin{equation}
	\left(F^{\cA\cA\cA}_\cA \right)_{\cA\cA} \cdot \ket{\mu}\otimes \ket{\mu} = \ket{\mu}\otimes \ket{\mu} \quad \text{ and } \quad R^{\cA\cA}_\cA \cdot \ket{\mu} =\ket{\mu}~,\label{algebra.str}
\end{equation}
which is consistent since $\cA$ has zero spin. 

Mathematically, $\ket{\mu} \in \mathrm{Hom}(\cA\otimes \cA,\cA)$ satisfying the above conditions define an associative commutative algebra structure on $\cA$ and is referred to as the \textit{multiplication} of the algebra~\cite{Kirillov:2001ti}.
A Lagrangian algebra object is an associative commutative algebra with a unique unit, that has trivial spin and satisfies the Lagrangian property \eqref{Lag.prop}. 
For more details, see Definition 1.1 and Figure 2 of~\cite{Kirillov:2001ti}. 
Here we have provided a geometric interpretation of these defining axioms in Figure \ref{fig:braidingfusionA}.

\subsubsection*{RCFT Interpretation}

Finally, we close this subsection with a discussion of what the topological boundary condition means for the boundary RCFTs.

For simplicity, we assume the bulk TQFT to be a Chern-Simons theory. 
The Chern-Simons theory has  a  standard Dirichlet boundary condition that supports the chiral WZW model.  
Let the space be  a disk with the standard Dirichlet boundary condition, and insert an anyon $a$ at a point in the bulk of the disk.  
This corresponds to the character $\chi_a$ of the boundary chiral algebra \cite{Moore:1989yh,Elitzur:1989nr}. 

Next, we make another hole in the disk with the conjugate boundary conditions.  This then gives rise to the diagonal modular invariant partition function of the boundary WZW model:
\ie
 Z[\tau,\bar\tau] = \sum_a \chi_a(\tau) \bar{\chi}_a(\bar \tau)~. 
 \fe
In other words, this is the same as  compactifying the  Chern-Simons theory  on an interval with the Dirichlet boundary condition for the gauge fields on the two boundaries of the interval.

Now let us further assume that the Chern-Simons theory admits a topological boundary condition. 
Then we can consider another configuration where on the second hole   we impose the topological boundary condition. 
This leads to a new modular invariant partition function 
\begin{equation} \label{modinv} Z[\tau]=\sum_aZ_{0a} \chi_a(\tau)~,\end{equation}
which is purely holomorphic. Indeed, the new topological boundary can be collapsed to a direct sum of anyons and hence introduces no new $\bar \tau$ dependence. 

More generally, we can put such a  Chern-Simons theory on a general Riemann surface times an interval. 
We impose the standard Dirichlet boundary condition on one end, and the topological boundary condition on the other end. 
The compactification of the interval then gives a holomorphic CFT. 
We  conclude that the existence of a topological boundary condition implies that the chiral algebra of the boundary RCFT can be extended to  a single module.\footnote{Note a subtlety regarding the modular invariance of~\eqref{modinv}: the invariance under the $S$ transformation is guaranteed by construction, but the $T$-matrix in 1+1d differs by $e^{2\pi i  \frac{c_-}{24}}$ from the $T$-matrix of the MTC and hence the invariance of~\eqref{modinv} under $T$ transformations is only guaranteed if $c_-=0 \mod{24}$ and not just $c_-=0 \mod{8}$ (\textit{cf.} \eqref{T.tilde}). Alternatively, we could always add some copies of $(E_8)_1$ to correct the issue.\label{footnote:TRT}} 

\subsection{Gauging the Lagrangian Algebra}\label{sec:gauging}
We have seen that the existence of a gapped boundary implies the existence of an anyon $\cA$ with various special structures and properties. As previously mentioned, such an anyon is called a Lagrangian algebra.

While $\cal A$ generally consists of non-Abelian anyons, it is possible to talk about the gauging/condensation of $\cal A$ and argue that gauging $\cA$ leads to a trivial theory. One can then recover the topological boundary condition by means of ``Dirichlet boundary conditions" after gauging $\cA$. 

Let us explain how this gauging is done geometrically. 
When we gauge a one-form symmetry (i.e.\ condense Abelian anyons) we are instructed to sum over all possible network of said anyons \cite{Gaiotto:2014kfa}. However, in the non-Abelian case it is no longer true that the only fusion channel of $\cal A$ and $\cal A$ is inside $\cal A$, and there is no subgroup (or more precisely, fusion subcategory) structure. Summing over general knots made out of $\cA$ then generally leads to contradictions because the crossing move is nontrivial. 

Instead of summing over all possible configurations of anyons, by gauging a Lagrangian algebra we will mean inserting a fine mesh of $\cA$.
A fine mesh is defined as the graph that is dual to a triangulation of the space-time manifold. 
More specifically, let $M_1$ be the 1-skeleton of a triangulation of a 3-manifold $M$, i.e.\ the union of the edges and vertices of the triangulation tetrahedra. 
Let $\widehat{M}_1$ be the dual 1-skeleton, i.e.\ the fine mesh.
Now consider fattening $M_1$ to get a regular neighborhood $N$ of $M_1$ which is topologically a handlebody. Moreover, deleting the interior of $N$ from $M$ we get $\widehat{N} = M - \mathrm{int}(N)$ which is also a handlebody, and which can be obtained by fattening the fine mesh $\widehat{M}_1$.\footnote{The decomposing of the 3-manifold $M$ into handlebodies $N$ and $\widehat{N}$ is known as a Heegaard splitting. See for instance \cite[Chapter 9]{rolfsen2003knots}.} Inserting the algebra anyon $\cal A$ on the fine mesh is equivalent to deleting $\widehat{N}$ from the space-time and putting the gapped boundary condition on the boundary of $N$.

Therefore after inserting $\cal A$ on the fine mesh, we are left with a handlebody $N$ of genus $g=e-v+1$ (where $e$ and $v$ are the number of edges and vertices of $M_1$) and with the gapped boundary condition on $\partial N$. To compute this partition function we can use the standard cutting and gluing similar to the one used in the context of 1+1d TQFTs \cite{Durhuus:1993cq}. Since we are assuming that the gapped boundary is simple, the Hilbert space on the disk $D^2$ is one-dimensional and the partition function is
\begin{equation}
	Z[N] = Z[D^3]^{1-g}~, 
\end{equation}
where $D^3$ is the three-dimensional ball. Since $Z[D^3]$ is positive by unitarity, we can set $Z[D^3]=1$ by adding the appropriate Euler counter-term on the boundary.\footnote{In the algebra language, this is equivalent to normalizing the multiplication such that composing the unit and counit give the quantum dimension of $\cal A$, i.e.\ $\beta_A=1$ in the notation of \cite{Fuchs:2002cm}.} Therefore, we see that after inserting a fine mesh of $\cA$ the partition function on any 3-manifold trivializes, which means that gauging a Lagrangian algebra anyon trivializes the theory.\footnote{Note that any associative commutative algebra with a unique unit and trivial spin in a unitary braided fusion category $\mathcal{B}$ can be gauged. The Lagrangian property \eqref{Lag.prop} only implies that after gauging all the nontrivial topological lines in $\mathcal{B}$ disappear.}

We can now define a topological interface between the original theory and the one obtained from gauging $\cA$. This interface is defined by putting the ``Dirichlet boundary conditions" for $\cA$, meaning that the mesh is terminated on the interface. More precisely, consider the original theory in region $M$ of the space-time that is connected via a codimension-one interface to region $M'$ where the gauged theory lives. To define the topological boundary condition, we first put the original theory on the whole space-time, i.e.\ on $M \cup M'$. We triangulate the space-time including the codimension-one interface. Now we can insert $\cA$ on the fine mesh in region $M'$, which is the graph dual to the tetrahedra of $M'$. Such a configuration defines a topological boundary condition for the original theory, since as we argued above the gauged theory is trivial.

The above discussion gives an unambiguous procedure to compute the partition functions of the gauged theory in terms of  those of the original theory decorated by the anyons. 
It is however not entirely clear to what extent it can really be interpreted as gauging some generalizations of  global symmetries. 
Among other things, there is no clear notion of a background gauge field for $\cal A$. 
We leave this point for future investigations.

\subsection{Topological Boundaries and the Turaev-Viro TQFT}\label{sec:TV}

In this subsection, we make contact with some known facts about the relation between the Lagrangian algebra, topological boundary conditions, and the Turaev-Viro TQFT.

For Abelian TQFT, Theorem \ref{thm:Lagsub} states that the existence of a topological boundary condition is equivalent to the existence of a Lagrangian subgroup.  
This theorem generalizes to non-Abelian TQFTs \cite{davydov2013witt,Fuchs:2012dt}: a bosonic TQFT has a  topological boundary condition if and only if it has a Lagrangian algebra $\cal A$. 
In Section \ref{sec:Lagalg} we gave a geometric interpretation of the only if part of the theorem.

What is the generalization of  Corollary \ref{cor:DW} that an Abelian TQFT admits a topological boundary if and only if it is an Abelian Dijkgraaf-Witten theory? 
It is  natural to ask if every non-Abelian TQFT with a topological boundary can be viewed as the pure gauge theory of something. 
The answer is given by the Turaev-Viro TQFT, which we briefly review below.

For any unitary fusion category $\cF$, there is a state-sum 2+1d TQFT known as Turaev-Viro(-Barret-Westbury) theory~\cite{Turaev:1992hq,Barrett:1993ab}.  
This class of TQFTs can be realized  as the low-energy limit of   the Levin-Wen string-net lattice model~\cite{Levin:2004mi,Lin:2020bak}.

 For instance, when the lines in the fusion category $\cF$ are all invertible ($\cF=\mathrm{Vec}_G^\omega$ for some $\omega \in H^3(G,U(1))$), the Turaev-Viro TQFT reduces to the discrete $G$-gauge theory with Dijkgraaf-Witten twist $\omega$~\cite{Dijkgraaf:1989pz}.
 
For more general unitary fusion category $\cF$, the corresponding Turaev-Viro TQFT can be thought of as $\cF$-gauge theory in the following sense. One begins with the trivial theory equipped with trivially acting topological surface defects with the fusion rules of $\cF$, and one then gauges these topological defects to obtain the Turaev-Viro TQFT~\cite{Carqueville:2018sld}.\footnote{The operation of gauging topological surface defects is the inverse operation of gauging topological line defects.}

Now we are ready to state another sufficient and necessary condition for the existence of a topological boundary condition: A TQFT admits  a topological boundary condition if and only if it is a Turaev-Viro TQFT \cite{Fuchs:2012dt, Freed:2020qfy}. 
Given that the latter can be viewed as a pure gauge theory of a unitary fusion category $\cal F$, this statement is the non-Abelian generalization of Corollary \ref{cor:DW}.

\subsubsection*{Summary on the Topological Boundary Conditions for Non-Abelian TQFTs}

Similar to the summary at the end of Section \ref{sec:completeObs} for the Abelian TQFTs, we summarize the sufficient and necessary conditions for topological boundary conditions for general non-Abelian TQFTs. 
If we work modulo invertible field theories (such as the $(E_8)_1$ CS theory), then the following conditions for a bosonic TQFT are equivalent:
\begin{itemize}
\item It admits a topological boundary condition.
\item It has a Lagrangian algebra.
\item It is a Turaev-Viro TQFT.\footnote{A Witten-Reshetikhin-Turaev type TQFT whose MTC is the Drinfeld center of a unitary fusion category is equivalent to a Turaev-Viro TQFT based on the same fusion category \cite{2010arXiv1004.1533K,2010arXiv1006.3501T}.}
\end{itemize} 
It would be interesting to find a complete list of obstructions similar to $\xi_n$ for the Abelian TQFTs .

\subsection{The Galois Action and the Higher Central Charges}\label{sec:nonAbhigherc}

 We have just seen that the existence of a Lagrangian algebra anyon $\cal A$ is necessary and sufficient condition for the existence of a gapped boundary. But in practice it is not easy to check whether such an $\cA$ exists in a given theory, since this requires knowledge of e.g. the $F$-matrices. 

We will now present some necessary conditions which do not require the $F$ or $R$ matrices, but only depend on the modular data, i.e. the $S$- and $T$-matrices.
In particular, we will discuss the non-Abelian generalizations of the  higher central charges $\xi_n$ which have been proved to be obstructions to topological boundary conditions \cite{Ng:2018ddj}.

One simple but very restrictive condition that we have already discussed is the existence of an integer-valued vector $ Z_{0a}$ satisfying
\begin{equation}
	SZ=Z~, \qquad TZ=Z~, \qquad Z_{00} = 1~.\label{eigenvector}
\end{equation}
The rest of this subsection will be dedicated to deriving some more subtle conditions. To introduce these conditions, some background information will be necessary. 

Say that we fix a basis of simple anyons $\mathcal{I}=\{ a, b, c, \dots \}$ with $N_{ab}^c\in \mathbb{Z}_{\geq 0}$. The space of MTC data consistent with this basis choice is given by the space of solutions to a set of polynomial equations, such as the pentagon and hexagon identities. This means that the $R,F,S,$ and $T$ matrices take values in a certain field extension of $\mathbb{Q}$. Associated to this field extension is a Galois group, and we can use Galois conjugation to map a given set of MTC data to a new set of MTC data~\cite{DeBoer:1990em,Coste:1993af} (see also~\cite{Buican:2019evc, Harvey:2019qzs}).

For the modular data in theories with vanishing chiral central charge $c_-=0$ mod 8, the relevant field extension and Galois group are~\cite{Bantay:2001ni,Ng:2012ty}
\begin{equation}
	\bQ(T,S) = \bQ\left(e^{\frac{2\pi i}{N_\mathrm{FS}}}\right)~, \qquad \Gal(T,S) = \mathbb{Z}_{N_\mathrm{FS}}^*~,
\end{equation}
where $\bZ_{N}^*$ is the multiplicative group consisting of all elements $n\in \bZ_N$ such that $\mathrm{gcd}(n,N)=1$.\footnote{If the chiral central charge does not vanish, the $S$ matrix may contain elements which are not in the above field extension.} 
Here $N_\mathrm{FS}$ is the Frobenius-Schur exponent, i.e.\ the smallest integer such that $\theta(a)^{N_\mathrm{FS}}=1$ for  of all anyons $a$.

Let $\sigma$ be an element of $\Gal(T,S)$. By abuse of notation, we will identify it with the corresponding integer number $\sigma$ mod $N_{\mathrm{FS}}$. Anytime we encounter an $N_\mathrm{FS}$-th root of unity $\zeta$, we simply raise it to the corresponding power to obtain its Galois conjugate,
\ie
\sigma\left(  \sum_{n=1}^{N_\mathrm{FS}} q_n \, \zeta^n \right) = \sum_{n=1}^{N_\mathrm{FS}} q_n \, \zeta^{\sigma n}~,\qquad \text{where } q_n \in \bQ~.
\fe

The $T$-matrix therefore transforms in an obvious way 
\begin{equation}
	\sigma(T) = T^\sigma
\end{equation}
since $\sigma(\theta_a) = \theta_a^\sigma$. 
The $S$-matrix transforms in a more complicated way \cite{Coste:1993af}.    There is a group homomorphism from the field extension into signed permutations   of the labels $\mathcal{I}$, such that the element $\sigma\in \mathbb{Z}_{N_\mathrm{FS}}^*$ maps to some permutation $\sigma:\mathcal{I} \to \mathcal{I}$ (by further abuse of notation) obeying
\begin{equation}
	\sigma(S_{ab}) = \epsilon_{\sigma}(a) S_{\sigma(a) b}  = \epsilon_{\sigma}(b) S_{a \sigma(b)} ~, \label{S.galois}
\end{equation}
with $\epsilon_\sigma(a)=\pm 1$. The fact that the Galois action on the $S$-matrix simply induces a (signed) permutation of the anyons is not obvious. Note that if we were to consider the ratios 
$S_{ab}/S_{a0}$ then the action of the Galois group would reduce to just a permutation of the $a$ anyon. 

Another important result is the congruence subgroup property of the modular representation defined by the $S$- and $T$-matrices~\cite{Bantay:2001ni,Ng:2012ty}. Note that in general the $S$- and $T$-matrices define a projective representation of the modular group because of the factor of $e^{2\pi i \frac{c_-}{8}}$ in \eqref{modular.rep}. However, when $c_- = 0 \mod{8}$ this factor is trivial and we get an ordinary representation $\rho$ of $\mathrm{SL}(2,\bZ)$ given by
\begin{equation}
	\rho: \left(
	\begin{array}{cc}
		0 & -1 \\
		1 & 0 \\
	\end{array}
	\right) \mapsto S~, \qquad \rho: \left(
	\begin{array}{cc}
		1 & 1 \\
		0 & 1 \\
	\end{array}
	\right) \mapsto T~.
\end{equation}
The congruence property then says that the kernel of this modular representation contains the principal congruence subgroup $\Gamma(N_\mathrm{FS})$, defined by
\begin{equation}
	\Gamma(N) = \left\{ \left(
	\begin{array}{cc}
		a & b \\
		c & d \\
	\end{array}
	\right) \in \mathrm{SL}(2,\bZ) : \left(
	\begin{array}{cc}
		a & b \\
		c & d \\
	\end{array}
	\right) = \left(
	\begin{array}{cc}
		1 & 0 \\
		0 & 1 \\
	\end{array}
	\right) \pmod{N} \right\}~.
\end{equation}
Note that $\Gamma(N)$ is just the kernel of the linear map $\pi_{N} : \mathrm{SL}(2,\bZ) \to \mathrm{SL}(2,\bZ_{N})$ defined by taking the mod $N$ reduction of the elements of $\mathrm{SL}(2,\bZ)$. Therefore the modular representation $\rho$ factors through $\pi_{N_\mathrm{FS}}$, and hence can be thought of as a representation of $\mathrm{SL}(2,\bZ_{N_\mathrm{FS}})$.\footnote{More precisely, there exist a representation $\rho_{N_\mathrm{FS}}$ of $\mathrm{SL}(2,\bZ_{N_\mathrm{FS}})$ such that $\rho = \rho_{N_\mathrm{FS}} \circ \pi_{N_\mathrm{FS}}$.} This allows us to compute the Galois conjugate modular representation $\rho_\sigma$ in terms of $\rho$, without knowing the signed permutation:
\begin{equation} 
	\rho_\sigma \left(
	\begin{array}{cc}
		a & b \\
		c & d \\
	\end{array}
	\right)  = \rho \left(
	\begin{array}{cc}
		a & \sigma b \\
		\bar{\sigma} c & d \\
	\end{array}
	\right)~, \label{Galois.congruence}
\end{equation}
where $\bar \sigma \in \mathbb{Z}$ is a multiplicative inverse of $\sigma$ modulo $N_\mathrm{FS}$, i.e.\ $\sigma \bar \sigma = 1 \pmod{N_\mathrm{FS}}$.

We can write the Galois action on the $S$- and $T$-matrices as $\sigma(S) = G_\sigma S = S G_\sigma^{-1}$ and $\sigma(T) = T^\sigma$, where $\left(G_\sigma\right)_{ab} = \epsilon_{\sigma}(a) \delta_{\sigma(a)b}$.  
Using \eqref{Galois.congruence}, we find an alternative expression for $\sigma(S)$ that doesn't involve $\epsilon_\sigma(a)$: 
\ie
\sigma(S) = T^\sigma S T^{\bar \sigma} S T^\sigma\,.\label{GaloisS}
\fe
It can further be shown that $\sigma^2(T)= G_\sigma T G_\sigma^{-1}$   \cite[Theorem II]{Ng:2012ty}.

\subsubsection*{Galois Action on Lagrangian Algebras}
If the original theory has a gapped boundary, then as we have discussed there exists a Lagrangian algebra anyon $\cA = \bigoplus_{a} Z_{0a} a$, and hence a non-negative integer eigenvector $Z$ satisfying $SZ=Z$. We will now show that upon Galois conjugation, the algebra anyon is left unchanged!

Indeed, we may begin by doing a Galois conjugation to both sides of the equation $SZ=Z$, giving
\begin{equation}
	\sigma(S_{ab}) Z_{0b} = Z_{0a}~.\label{galois.conj}
\end{equation}
Here we have used the fact that the $Z_{0a}$ are integers, and thus are left invariant by Galois conjugation.
Note that \eqref{galois.conj} is equivalent to $\epsilon_\sigma(a) S_{\sigma(a)b}Z_{0b} = Z_{0a}$. Using $SZ=Z$ once again, we can then derive a constraint
\begin{equation}
	Z_{0\sigma(a)} = \epsilon_\sigma(a) Z_{0a}~.\label{Z.galois}
\end{equation}
Since the algebra anyon is given by $\cA = \bigoplus_{a} Z_{0a} a$, this means that the permutation can only send anyons in $\cA$ to other anyons in $\cA$ with the same multiplicity. In addition, since $Z$ is positive, $\epsilon_\sigma(a)=1$ when $a$ is contained in $\cA$, and hence we conclude that 
\ie
Z_{0\sigma(a)} = Z_{0a} \qquad \Rightarrow \qquad \sigma(\cA)=\cA~.
\fe
In other words, the permutation induced by the Galois group is constrained in such a way that it preserves the algebra anyon!

The various nice properties of $\cA$ are also preserved under Galois conjugation. For example, since under Galois conjugation the spins are simply multiplied by $\sigma$, and since the original spins vanished for $\cA$, the same is also true after the conjugation, i.e.\ $\sigma(T)Z=Z$. We have also seen in (\ref{galois.conj}) that $\sigma(S)Z=Z$. 
Finally, since for the algebra anyon the $F$ and $R$ matrices are ``trivial", the same will remain true after conjugation. More precisely, we must show that for the Galois conjugate theory there exists a $\ket{\mu} \in V_\cA^{\cA\cA}$ satisfying \eqref{algebra.str}. To show this, first note that since the fusion coefficients do not change under Galois conjugation, we can identify the fusion vector spaces $V_c^{ab}$ before and after Galois conjugation. The multiplication of the algebra is just a particular vector
\ie
\ket{\mu} \in V_\cA^{\cA\cA}=\bigoplus_{a,b,c \in I} \bigoplus_{\alpha=1}^{Z_{0a}} \bigoplus_{\beta=1}^{Z_{0b}} \bigoplus_{\gamma=1}^{Z_{0c}}  V_c^{ab}~.
\fe
In this fixed basis \eqref{algebra.str} is equivalent to
\begin{equation}
	\bra{\nu \otimes \rho} \left(F^{\cA\cA\cA}_\cA \right)_{\cA\cA} \ket{\mu \otimes \mu} = \delta_{\nu\mu}\delta_{\rho\mu} \quad \text{ and } \quad \bra{\nu} R^{\cA\cA}_\cA \ket{\mu} = \delta_{\nu\mu}~,\label{F.and.R}
\end{equation}
for all $\ket{\nu},\ket{\rho} \in V_\cA^{\cA\cA}$. Since the numbers on the RHS of equations in \eqref{F.and.R} are integer, they do not change under Galois conjugation. Hence the algebra multiplication $\ket{\mu}$ is preserved under Galois conjugation.

\subsubsection*{Obstructions from Galois Conjugation}
Having shown that Galois conjugation maps a theory with a Lagrangian algebra to another theory with such an algebra, we are ready to draw some conclusions regarding gapped boundaries. 

We start from a TQFT that admits a topological boundary condition. 
Since its (possibly non-unitary) Galois conjugate theory  also admits  a topological boundary, we  expect that similar to the unitary case its $S^3$ partition function 
\ie
Z_\sigma[S^3] = \sigma(S_{00})
\fe
is positive. Here $Z_\sigma[S^3]$ stands for the $S^3$ partition function of the Galois conjugate theory.
Indeed, $\sigma(S_{00})$ is positive. 
The proof is given as follows: noting that $\sigma(S_{00}) = \epsilon_\sigma(0)S_{\sigma(0)0}$ and using $Z_{00}\neq 0$ as well as the constraint~\eqref{Z.galois} to find $\epsilon_\sigma(0)=1$, we find 
\begin{equation}\label{ConjthreeS}
	Z_\sigma[S^3] = S_{\sigma(0)0}
\end{equation}
which is  positive.

We can now obtain some necessary conditions for the existence of gapped boundaries in our original theory by computing the $S^3$ partition function of its Galois conjugate using a different surgery presentation. In particular, $S^3$ is homeomorphic to the lens space $L(1,1)$, and therefore (\textit{cf.} equation \eqref{lens})
\begin{equation*}
	Z_\sigma[S^3] = \sigma(STS)_{00}~.
\end{equation*}
Of course if the original theory has a gapped boundary then it has a vanishing chiral central charge and hence $(STS)_{00}=S_{00}$, which is an identical expression to what we wrote in~\eqref{ConjthreeS}.

Using \eqref{Galois.congruence} we find
\begin{equation*}
	STS = \rho\left(
	\begin{array}{cc}
		-1 & 0 \\
		1 & -1 \\
	\end{array}
	\right) \quad \Rightarrow  \quad \sigma(STS) = \rho\left(
	\begin{array}{cc}
		-1 & 0 \\
		\bar{\sigma} & -1 \\
	\end{array}
	\right) = S T^{\bar \sigma} S~.
\end{equation*}
Thus the $S^3$ partition function of the Galois conjugate theory is $(S T^{\bar \sigma} S)_{00}$, which is just the $L(\bar \sigma,1)$ partition function of the original theory. This gives us a prediction for the partition functions on lens spaces: 
\begin{equation}
	Z\big[ L(\sigma,1) \big] = S_{\bar{\sigma}(0)0} \geq S_{00}~,\label{S3.Galois}
\end{equation}
which in particular is positive. 
Therefore we find that if a theory has a gapped boundary then its lens space partition function, which admits the simple formula
\begin{equation}
	Z\big[ L(\sigma,1) \big] = \frac{1}{\cD^2} \sum_a d_a^2 \theta_a^\sigma~,
\end{equation}
must be positive when $\mathrm{gcd}(\sigma,N_\mathrm{FS})=1$. 

 To summarize,  the higher central charges for general non-Abelian TQFTs are defined as
\ie
\xi_\sigma := { \sum_a d_a^2 \, \theta(a)^\sigma \over| \sum_a d_a^2 \, \theta(a)^\sigma|} \,.
\fe
A TQFT admits a topological boundary condition only if $\xi_\sigma=1$ for all $\mathrm{gcd}(\sigma,N_\mathrm{FS})=1$ \cite{Ng:2018ddj}.

Moreover, the inequality on the RHS of \eqref{S3.Galois} implies the following interesting number theoretic property of $\cD = \sqrt{\sum_a d_a^2}$,
\begin{equation}
	 \cD \geq \sigma(\cD) > 0~.
\end{equation}
This means that all roots of the minimal polynomial of $\cD$ are positive and less than or equal to $\cD$.

\subsubsection*{RCFT Interpretation via Galois Conjugation}

We close by briefly mentioning the relation of these obstructions to RCFT. The Galois action defined above on the modular data $S$ and $T$ extends in the straightforward way to 1+1 dimensions.  At the level of the RCFT characters, it was shown in \cite{Harvey:2018rdc,Harvey:2019qzs} that the Galois transformation given by $\sigma\in \mathbb{Z}_{N_\mathrm{FS}}^*$ induces the action of the Hecke operator $\Tt_{\bar{\sigma}}$, which is defined as follows. Consider an RCFT with $d$ characters, which may be organized into a $d$-dimensional vector-valued modular function $\mathbf{\chi} = (\chi_1, \dots, \chi_d)$. Then the $\sigma$-th Hecke operator $\Tt_\sigma$ for any prime $\sigma$ such that  $\mathrm{gcd}(\sigma, \widetilde{N}_\mathrm{FS})=1$ is defined as\footnote{Note that $\widetilde{N}_\mathrm{FS}$ is the order of the RCFT $T$-matrix, which is related to the TQFT $T$-matrix used so far by $\widetilde{T} = e^{-2 \pi i \,{c_-\over 24}} T$ (\textit{cf.} \eqref{T.tilde}). By \cite[Theorem II]{Ng:2012ty} we have $N_\mathrm{FS} \mid \widetilde{N}_\mathrm{FS} \mid 12N_\mathrm{FS}$.
}
\begin{equation}
	\left( \Tt_\sigma \chi \right)_a(\tau) = \epsilon_{\bar\sigma}(a) \chi_{\bar\sigma(a)}(\sigma \tau) + \sum_{j=0}^{\sigma-1} \chi_a\left(  {\tau + j \widetilde{N}_\mathrm{FS} \over \sigma}\right)~.
\end{equation}
Having the expression of $\Tt_\sigma$ for prime $\sigma$, Hecke operators for $\sigma$ coprime to $\widetilde{N}_\mathrm{FS}$ but not necessarily prime are constructed in appendix of \cite{Harvey:2018rdc}. 

The action of the Hecke operator on the characters gives a new set of characters, which may be interpreted as those of the Galois conjugate RCFT. An illustrative example is to consider the action of $\Tt_\sigma$ on the characters of the Lee-Yang minimal model; one finds that the action of $\Tt_7, \Tt_{13},$ and $\Tt_{19}$ gives rise to the respective characters of $(G_2)_1$, $(F_4)_1$, and $(E_{7 {1\over 2}})_1$, all of which are in the same Galois orbit. 

With this in mind, we may reinterpret the phases of $Z[L(\sigma,1) ]$ discussed above as the usual chiral central charges of the 1+1d RCFTs related by the Hecke transformation $\Tt_\sigma$ to the original RCFT on the boundary. 
In other words, the higher central charges are just the usual chiral central charges for appropriate conjugate CFTs.

For Abelian TQFTs with a one-form symmetry group $G$, we had a larger set of obstructions in addition to the higher central charges.
These additional invariants arise from  three-manifolds with $\text{gcd}( |H_1(M)|  , |G|)=1$ (see Section \ref{sec:main}).
Furthermore, they also arise from  lens spaces with an extended range of allowed $n$, i.e., those $n$ such that ${\rm gcd}\left( n, {2 |G| \over {\rm gcd}(n,2|G|)}\right) =1$ (see Section \ref{sec:completeObs}). 
We do not have a concise 1+1d interpretation for these higher obstructions.

\section*{Acknowledgements}

We thank M. Barkeshli, N. Benjamin, F. Burnell, Y.-A. Chen, T. Devakul, D. Freed, P. Gorantla, H. T. Lam, N. Seiberg, C. Teleman, J. Wang, and F. Yan  for helpful discussions. 
We also thank G. Moore for comments on the draft. 
ZK is supported in part by the Simons Foundation grant 488657 (Simons Collaboration on the Non-Perturbative Bootstrap) and the BSF grant no. 2018204.
SHS was also supported by the Simons Collaboration on Ultra-Quantum Matter, which is a grant from the Simons Foundation (651440, NS).

\appendix

\section{Review of 2+1d TQFTs}\label{app:3d.TFT}

In this appendix we review the classification of unitary 2+1d TQFTs. These theories are described by a pair
\begin{equation}
	(\cC, c_-) \label{tqft.data}
\end{equation}
where $\cC$ is a \emph{unitary modular tensor category} (UMTC) \cite{Moore:1988qv, Turaev+2010, bakalov2001lectures, Kitaev:2005hzj, wen2016theory} and $c_- \in \bQ$ is the chiral central charge\footnote{$c_-=c-\bar{c}$ is related to the thermal Hall conductance $K_H=c_-\frac{\pi k_B^2}{6\hbar}T$.} of the boundary CFT. Note that $\cC$ determines the chiral central charge $c_-$ modulo $8$, and hence $\cC$ and $c_-$ are not independent. TQFTs defined by such data are known as Witten-Reshetikhin-Turaev theories~\cite{Witten:1988hf,Reshetikhin:1991tc}.

When $c_-\neq0$ the theory is not strictly speaking topological and the partition function has a mild dependence on the geometry/metric.\footnote{Note that the scheme that we are using here is different from Witten's original quantization of Chern-Simons theory in \cite{Witten:1988hf}. Witten adds a (not necessarily properly quantized) gravitational Chern-Simons term to cancel the metric dependent of the theory and instead introduces a framing dependence. In our scheme the partition function does not depend on the framing of the 3-manifold. } The metric dependence implies a perturbative gravitational/diffeomorphism anomaly \cite{AlvarezGaume:1983ig} on the boundary. So for $c_-\neq0$ the theory on the boundary cannot be gapped. Since in this paper we are interested in 2+1d theories with a gapped boundary, from now on we only focus on theories with zero chiral central charge, i.e.\ $c_-=0$.

In this work we are only interested in theories with a unique vacuum on $S^2$. Note that vacuum degeneracy on $S^2$ leads to nontrivial topological point operators via the state/operator correspondence. In a unitary theory the algebra of topological point operators is a semisimple Frobenius algebra which has a complete set of idempotents (projection operators). These projection operators correspond to different superselection sectors (universes) of the theory~\cite{Hellerman:2006zs,Tanizaki:2019rbk,Komargodski:2020mxz}. Hence any unitary 2+1d TQFT can be decomposed into theories with a one-dimensional $S^2$ Hilbert space. So there is no loss of generality by restricting to theories with a unique vacuum on $S^2$.

The line defects (a.k.a anyons) of a 2+1d TQFT are an important set of observables that determine the theory uniquely. These line defects along with their various properties form the unitary modular tensor category $\cC$. Mathematically speaking, a category consists of a set of objects and a set of morphisms between those objects.  The objects of the category $\cC$ correspond to the line defects of the TQFT, while the morphisms between two objects correspond to junctions between the corresponding lines. A unitary modular tensor category has extra structures besides the objects and morphisms. Below we review these structures in terms of the line defects of the TQFT. For more details see Appendix E of \cite{Kitaev:2005hzj} or Section 5 of \cite{Benini:2018reh}. 

\begin{enumerate}
	\item \textbf{Fusion:}
	\begin{equation}
		a \otimes b \simeq \bigoplus_{c \in I} N_{ab}^c \, c~.
	\end{equation}
	In a general fusion category, multiplication does not have to be commutative. However, in a UMTC, due to the existence of an invertible $R$-matrix (below) it is true that $N_{ab}^c=N_{ba}^c$. $N_{ab}^c$ is interpreted as the dimension of the Hilbert space of the junction of three lines. There are two related vector spaces, $V^{ab}_c$ and $V_{ab}^c$, corresponding to the splitting and fusion vector spaces, respectively. They satisfy ${\rm dim}(V^{ab}_c)={\rm dim}(V_{ab}^c)=N_{ab}^c$. 
	Denoting the trivial anyon by $0$ we can regard any anyon as itself being a vector space $V_{a0}^a$, seen by attaching the trivial anyon. 
	
More generally there are vector spaces $V_{a_1...a_n}^{b_1...b_m}$ corresponding to the fusion space of $n$ anyons into $m$ anyons. These vector spaces have a natural basis in terms of tensor products of the elementary spaces $V^{ab}_c$ and $V_{ab}^c$. For instance, we may obtain any vector in $V_{ab}^{cd}$ by first picking a vector in $V_{ab}^e$ for some $e$, and then another vector in $V^{cd}_e$. Therefore 
\ie
V^{cd}_{ab} = \bigoplus_e (V^{cd}_e\otimes V_{ab}^e)~.
\fe

For every anyon $a$ there is an anyon $\bar a$ with the property that $V_{a\bar a}^0$ and $V_{0}^{a\bar a}$  are one-dimensional vector spaces. In addition, there are canonical isomorphisms from e.g. $V_{c}^{ab}$ to $V_{\bar a c }^b$, given by taking the $a$ anyon in $V_{c}^{ab}$ backwards in time. Requiring that these isomorphisms are unitary gives a preferred normalization for the inner product on the splitting and fusion spaces. 

In this normalization, which is widely used in the literature, an unknot of $a$ gives
\begin{equation}
	d_a = \raisebox{-1.1em}{\begin{tikzpicture}
			\draw [thick, decoration = {markings, mark=at position .33 with {\arrowreversed[scale=1.5,rotate=10]{stealth}}}, postaction=decorate] (0,0) ++(50:.5) arc (50:410:.5);
			\node at (-.8,0) {\small $a$};
	\end{tikzpicture}}~, \label{qdim}
\end{equation}
which is called the \emph{quantum dimension} of $a$. By bending the lines and rotating the figure by 90 degrees one finds an isomorphism between $V_{ab}^{cd}$ and $V_{a\bar c}^{\bar b d}$, and further using the decomposition to splitting and fusion spaces we find 
\ie
\sum_e N_{ba}^eN_{cd}^e = \sum_f N_{a\bar c}^f N_{\bar b d}^f~.
\fe
Using that $N_{a\bar c}^f=N_{fc}^a=N_{cf}^a$ and $ N_{\bar b d}^f=N_{bf}^d$,  we rewrite the above equation as 
\ie
\sum_e N_{ba}^eN_{cd}^e = \sum_f N_{ bf}^dN_{ cf}^a ~.
\fe
Defining the matrices ${\bf N_a}=(N_a)^b_c$, the above equation reads ${\bf N_b}{\bf N_c}^T={\bf N_c}^T{\bf N_b}$, i.e.\ $[{\bf N_b},{\bf N_c}^T]=0$ for all $b,c$. This guarantees that the matrices ${\bf N_c}$ are all mutually diagonalizable. An important fact is that the vector of quantum dimensions $d_a$ satisfies 
\begin{equation}\label{FSV} d_ad_b=\sum_c N_{ab}^cd_c~,\end{equation}
i.e.\ ${\bf d}$ is an eigenvector of ${\bf N_a}$ with eigenvalue $d_a$.

Above we have used the fact that vector spaces $V_{ab}^{cd}$ and $V_{a\bar c}^{\bar b d}$ have the same dimension. More generally, the vector space $V_d^{abc}$ admits distinct decompositions in terms of splitting and fusion spaces:
\ie
\bigoplus_e V_{e}^{ab}\otimes V_{d}^{ec}~,\quad \bigoplus_f V_{d}^{af}\otimes V_{f}^{bc}~.
\fe
In addition to the requirement that these lead to a vector space of the same dimension, there exist fusion matrices
$F^{abc}_d:\bigoplus_e V_{e}^{ab}\otimes V_{d}^{ec} \to \bigoplus_f V_{d}^{af}\otimes V_{f}^{bc}$
which allow us to translate all vectors in one basis to the other. The $F$-matrices (6j-symbols)
\begin{equation}
	\left(F^{abc}_d\right)_{ef}: V_{e}^{ab}\otimes V_{d}^{ec} \to V_{d}^{af}\otimes V_{f}^{bc} \label{f-matrices}
\end{equation}
are famously subject to the pentagon equation and also the triangle equation (the latter simply ensures that no harm is done in attaching the trivial anyon at will).

	\item \textbf{R-matrices (Braiding):}
	
Above we have already used the fact that $N^{c}_{ab}=N^c_{ba}$. In fact, there is an invertible linear map between the vector spaces $V^{ab}_c$ and $V_{c}^{ba}$:
\ie
R^{ab}_c:V^{ab}_c\longrightarrow V_{c}^{ba}
\fe which is pictorially represented as
\be
{\begin{tikzpicture}[baseline=0]
\draw [thick, decoration = {markings, mark=at position .7 with {\arrow[scale=1.5]{stealth}}}, postaction=decorate] (0,-.5) node[below] {\small $c$} to (0,0) node[right] {\footnotesize $\mu$};
\draw [thick] (0,0) arc [radius=.3, start angle=240, end angle=120];
\draw [thick, decoration = {markings, mark=at position .75 with {\arrow[scale=1.5]{stealth}}}, postaction=decorate] (0,0.52) to +(30:.5) node[right] {\small $b$};
\draw [thick] (0,0) arc [radius=.3, start angle=-60, end angle=42];
\draw [thick, decoration = {markings, mark=at position .8 with {\arrow[scale=1.5]{stealth}}}, postaction=decorate] (0,0.52) ++(150:.1) to +(150:.4) node[left] {\small $a$};
\end{tikzpicture}}
 = \hspace{0.1 in} \sum_\nu (R^{ab}_c)_{\mu\nu} 
{\begin{tikzpicture}[baseline=-10]
\draw [thick, decoration = {markings, mark=at position .5 with {\arrow[scale=1.5]{stealth}}}, postaction=decorate] (0,-.8) node[below] {\small $c$} to (0,0);
\draw [thick, decoration = {markings, mark=at position .7 with {\arrow[scale=1.5]{stealth}}}, postaction=decorate] (0,0) node[below right] {\footnotesize $\nu$} to (-.6,.5) node[left] {\small $a$};
\draw [thick, decoration = {markings, mark=at position .7 with {\arrow[scale=1.5]{stealth}}}, postaction=decorate] (0,0)--(.6,.5) node[right] {\small $b$};
\end{tikzpicture}}~. \label{R.matrix}
\ee

 We can write the matrix in components as $R^{ab}_{c; \mu \nu}$.
The $R$-matrix is subject to two hexagon equations that ensure its compatibility with the $F$ matrices. An important conceptual point is that braiding an anyon $a$ around an anyon $b$ depends on their fusion channel $c$, and not only on the type of the anyons $a,b$ alone. 
Using the $R$-matrix it is possible to define the \emph{topological spin} as 
\begin{equation}
	\theta(a)={1\over d_a} \sum_c d_c \Tr(R^{aa}_c)= \frac1{d_a} \;\; \raisebox{-1.5em}{\begin{tikzpicture}
			\draw [thick] (0,0) ++(45:.5) arc (45:315:.5);
			\draw [thick] (0,0) ++(-45:.5) to +(45:1);
			\draw [thick, decoration = {markings, mark=at position .46 with {\arrowreversed[scale=1.5,rotate=10]{stealth}}}, postaction=decorate] (0,0) ++(-45:1) ++(45:.5) arc (-135:135:.5);
			\node at (2.20,0) {\small $a$};
			\draw [thick] (0,0) ++(45:.5) to +(-45:.3); \draw [thick] (0,0) ++(45:.5) ++ (-45:.7) to +(-45:.3);
	\end{tikzpicture}}
	~. \label{tspin}
\end{equation}
The topological spin is always a root of unity with $\theta(a)=\theta({\bar a})$ \cite{Vafa:1988ag}. We sometimes write $\theta(a)=e^{2\pi i s_a}$.

For our purposes it is important to quote the following nontrivial property of the braiding matrix $R$. Applying the matrix twice gives an automorphism of the splitting space 
\ie
R^{ba}_cR^{ab}_c:V^{ab}_c\longrightarrow V_{c}^{ab}~,
\fe
and it turns out that this map is proportional to the identity map up to an overall phase given in terms of the topological spins:
\begin{equation}\label{twiceR}
R^{ba}_cR^{ab}_c={\theta(c)\over \theta(a)\theta(b)}id_{V_c^{ab}}~.
\end{equation}

The topological spins are constrained by the hexagon identities through their relation to the $R$-matrix. It is possible to derive one general constraint that is independent of the $F$-matrices and is given purely in terms of the the topological spins and the dimensions $N_{ab}^c$ of the fusion spaces. For arbitrary labels $w,x,y,z$ this constraint is 
\begin{equation}\label{thetac}\prod_p \theta(p)^{N_{xy}^pN^{\bar p}_{wz}+N^p_{xz}N^{\bar p}_{wy}+N_{yz}^pN_{wx}^{\bar p}}=\left(\theta(w)\theta(x)\theta(y)\theta(z)\right)^{\sum_q N_{xy}^qN_{wz}^{\bar q}}~.\end{equation}

	\item \textbf{Modular $S$ and $T$ Matrices:}
	
From the above data, it is possible to construct $S$ and $T$ matrices which behave similarly to the $S$ and $T$ matrices of two-dimensional conformal field theory. The components $S_{ab}$ of the S-matrix are defined by 
\begin{equation}\label{Smat} S_{ab}\,\,=\,\,{1\over{\cal D}}\sum_c N_{a\bar b}^c {\theta(c)\over \theta(a)\theta(b)}d_c \,\,=\,\, {1\over{\cal D}} \raisebox{-1.1em}{\begin{tikzpicture}
\draw [thick, decoration = {markings, mark=at position .33 with {\arrowreversed[scale=1.5,rotate=10]{stealth}}}, postaction=decorate] (0,0) ++(60:.5) arc (60:390:.5);
\node at (-.8,0) {\small $a$};
\draw [thick, decoration = {markings, mark=at position .88 with {\arrowreversed[scale=1.5,rotate=10]{stealth}}}, postaction=decorate] (.7,0) ++(-120:.5) arc (-120:210:.5);
\node at (-.05,0) {\small $b$};
\end{tikzpicture}}
\end{equation}
where
\begin{equation}
	{\cal D}=\sqrt{\sum_ad_a^2}=S_{00}^{-1}~.\label{eq:totD}
\end{equation}
is the \emph{total quantum dimension} of $\cC$.\footnote{In the mathematical literature, $\mathrm{dim}(\cC)=\sum_a d_a^2$ is called the (global) dimension of $\cC$.} We will also find the following result useful, 
\be
{\begin{tikzpicture}[baseline=-5]
\draw [thick, decoration = {markings, mark=at position .3 with {\arrowreversed[scale=1.5,rotate=15]{stealth}}}, postaction=decorate] (0,0) ++(100:.5 and .3) arc (100:440:.5 and .3);
\node at (-.7,0) {$a$};
\draw [thick] (0,-.8) to (0,-.4); \draw[thick, decoration = {markings, mark=at position .5 with {\arrow[scale=1.5,rotate=0]{stealth}}}, postaction=decorate]  (0,-.2) to (0,.6) node[above] {$b$};
\end{tikzpicture}}
\hspace{0.1 in}= \hspace{0.1 in} \frac{S_{ab}}{S_{0b}} \quad
{\begin{tikzpicture}[baseline=-5]
\draw [thick, decoration = {markings, mark=at position .7 with {\arrow[scale=1.5]{stealth}}}, postaction=decorate] (0,-.8) to (0,.6) node[above] {$b$};
\end{tikzpicture}} ~.
\label{eq:defofS}
\ee

$S$ is a symmetric matrix and, in addition, $S_{\bar a\bar b} = S_{ab}$ and $S_{\bar a b}=S_{ab}^*$. Note that $S_{0c}={1\over{\cal D}} d_c$ and in the same way that $d_c$ was an eigenvector of $\bf N$, $S_{xc}$ is an eigenvector for every $x$ 
\begin{equation}\label{Seigen}{S_{ax}S_{bx}\over S_{0x}} =\sum_c N^c_{ab} S_{xc}~.\end{equation}
A few additional definitions: $C=\delta_{\bar a b}$, $T_{ab}=\theta_a\delta_{ab}$, and very importantly, 
\begin{equation}\label{ccc} e^{2\pi i c_-\over 8} = {1\over \cal D}\sum_a d_a^2\theta_a~.\end{equation} 
It requires a proof that the right hand side of~\eqref{ccc} is a pure phase -- we do not review the general proof here, but we will prove the Abelian case later.
The quantity $c_-$ is identified with the chiral central charge of the edge modes. The UMTC only determines it mod 8, which is an important fact we will come back to later.

These matrices satisfy\footnote{In a non-unitary theory there is an ambiguity in the sign of the $S$-matrix since there is no positivity constraint, hence \eqref{modular.rep} only determines $c_- \,\,\,{\rm mod}\,\,{4}$. More precisely if $(S,T,c_-)$ is a solution, $(-S,T,4+c_-)$ is also a solution.}
	\begin{equation}
		S^2=C~,\quad (ST)^3=e^{2\pi i c_- \over 8}C~,\quad C^2=1~.\label{modular.rep}
	\end{equation}
	 Note that it is possible to define $\widetilde T=\theta_ae^{-{2\pi i c_-\over 24}}\delta_{ab}$ in terms of which we get 
\begin{equation}
	S^2=C~,\quad (S\widetilde T)^3=C~,\quad C^2=1~. \label{T.tilde}
\end{equation}
Here the matrix $\widetilde T$ is  the $T$-matrix in the RCFT.

\end{enumerate}

For the classification of low-rank UMTCs see \cite{Rowell:2007dge, wen2016theory}.

\subsection*{The Abelian Case}
An anyon $a$ is called Abelian if $d_a=1$. This is equivalent to the requirement that for any other anyon $x$, $N_{xa}^c$ is non-vanishing for only one $c$.\footnote{The proof involves using~\eqref{FSV} along with the symmetries of the symbol $N_{ab}^c$. First we assume by contradiction that $N_{xa}^c\neq 0 $ and $N_{xa}^{d}\neq 0$ for $c\neq d$. Then we find $d_x\geq d_c+d_d$. But on the other hand $N_{xa}^c=N_{\bar a c}^x$, and therefore $d_c\geq d_x$. These two are clearly incompatible if the anyon $d$ exists.  } If all the anyons are Abelian the fusion rules therefore lead to an Abelian group $G$, and the corresponding theories are called Abelian TQFTs. The inverse of the anyon $a$ is $\bar a$. 

Next we must find $\theta(a)$ which is a map from the group to $U(1)$,
\ie
\theta: G\to U(1)~.
\fe
The main constraint on $\theta$ comes from~\eqref{thetac} if we plug $w=\bar x$ and $z=\bar y$. In this case we find
\begin{equation}\label{spinAbelian}\theta_{xy}\theta_{x\bar y}=\theta_x^2\theta_y^2~.\end{equation}
A neat way to encapsulate the information in~\eqref{spinAbelian} is to define a map
\ie 
B: G\times G \to U(1)
\fe
such that 
\begin{equation}\label{braidB}B(x,y)={\theta(xy)\over \theta(x)\theta(y)}~.\end{equation} $B$ is of course just the full braiding of anyon $x$ around anyon $y$, c.f. \eqref{twiceR}. In the special case of Abelian theories, the fusion channel of $x,y$ is unique and hence the braiding phase can be defined as a function of $x,y$ only. 
From property~\eqref{spinAbelian} we immediately conclude that 
$B(\bar x,y)=B^*(x,y)={1\over B(x,y)}$.
In fact, $B$ is a bilinear map on the Abelian group $G$ as it satisfies 
\begin{equation}\label{bilinear}B(zx,y)=B(z,y)B(x,y)~.\end{equation}
The condition $B(\bar x,y)=B^*(x,y)={1\over B(x,y)}$ follows from the bilinear property since $1=B(0,y)=B(x\bar x,y)=B(x,y)B(\bar x,y)$ and hence $B(\bar x,y)=B^*(x,y)={1\over B(x,y)}$. The topological spins $\theta$ are called a quadratic refinement of the bilinear map.

To derive the bilinear property~\eqref{bilinear},  we must show that $\theta(zxy)={\theta(zy)\theta(xy)\theta(zx)\over \theta(x)\theta(y)\theta(z)}$. This in turn follows from the fact that the $S$-matrix columns are eigenvectors of the fusion matrix, as per \eqref{Seigen}. 
Indeed, using \eqref{spinAbelian} we find $S_{xy}={1\over \sqrt {|G|}} {\theta(x\bar y) \over \theta(x)\theta(y)}={1\over \sqrt{|G|}} {\theta(x)\theta(y) \over \theta({xy})}$, and then \eqref{Seigen} reads
\ie
{\theta(a)\theta^2(x)\theta(b)\over \theta({ax})\theta({bx})} = {\theta(x)\theta({ab}) \over \theta({xab})}
\fe
which is exactly the required condition for~\eqref{bilinear} to hold.

Finally, the bilinear map $B(x,y)$ is non-degenerate in theories with a unitary $S$-matrix. That is, for every anyon, there is another anyon that braids with it nontrivially. 
From the unitarity of the $S$-matrix we have \begin{equation}\label{Ortho} |G|\delta_{a\bar b}= \sum_c B(a,c)B(b,c) ~.\end{equation}
Therefore if such an anyon $x_0\neq 0$ that has trivial braiding with all anyons existed we could have set 
$a=x_0$, $b=0$ and found $ |G| \delta_{x_0,0}= \sum_c B(x_0,c) $ which is a contradiction since the left hand side vanishes for   $x_0\neq 0$ while the right hand side is equal to $|G|$ by assumption.
(This statement holds very generally: The $S$-matrix~\eqref{Smat} is unitary when the braiding is non-degenerate.)

A simple consequence of~\eqref{Ortho} is $|G|\delta_{a 0}= \sum_c B(a,c)=\sum_c{\theta(a)\theta(c)\over \theta({ac})}$. For $a\neq 0$ we get $0=\sum_c{\theta(c)\over \theta({ac})}$ and hence also $0=\sum_{c,a\neq 0}{\theta(c)\over \theta({ac})}$. The sum over $a$ can be traded for a sum over all elements $b$ of $G$ such that $b=ac$ except the elements $b=c$ which can be added by hand. Hence,
$|G|=\sum_{c,b}{\theta(c)\over \theta({b})}=\sum_{c,b}\theta(c)\theta(b)^*$. This is equivalent to the statement that the right hand side of~\eqref{ccc} is a pure phase.

\section{Surgery Calculation of Partition Functions}\label{app:RT}

In this appendix we review the computation of 2+1d TQFT partition functions via the Dehn surgery presentation of 3-manifolds. Given a UMTC $\cC$, there is a topological invariant associated to any 3-manifold known as the Reshetikhin-Turaev (RT) invariant \cite{Reshetikhin:1991tc, Turaev+2010, bakalov2001lectures}. For $\cC$ with chiral central charge $c_- = 0 \pmod{8}$, the RT-invariant coincides with the partition function of metric-independent 2+1d TQFTs in question. 
Note that when $\cC$ is the category of representations of the quantum group associated with the Lie group $G$ at level $k$, the RT-invariant is almost the same as Witten's $G_k$ CS invariant \cite{Witten:1988hf}. The only difference is that in Witten's quantization the partition function of the theory depends on a choice of 2-framing \cite{atiyah1990framings} of the 3-manifold, arising from the addition of an improperly quantized gravitational CS term to cancel the metric dependence of the theory. In \cite{atiyah1990framings}, Atiyah showed that one can always define a canonical 2-framing, and it turns out that the partition function of Witten's CS theory on 3-manifolds with this canonical 2-framing coincides with the RT invariant.

\subsection*{Dehn Surgery}

In order to describe the computation of the RT invariant, we must first review the surgery presentation of 3-manifolds; for more details see Chapter 9 of the book by Rolfsen \cite{rolfsen2003knots}. 

\paragraph{Framed Links (Ribbon Graphs):} According to the Lickorish-Wallace theorem, any closed oriented 3-manifold can be obtained by performing Dehn surgery on a framed link (i.e. ribbon graph) in the 3-sphere $S^3$.  A link $L$ in $S^3$
\ie
L = K_1 \cup K_2 \cup \cdots \cup K_m 
\fe
 is a collection of knots $K_1 , K_2, \cdots, K_m \subset S^3$ that might link non-trivially with each other. Each knot $K_i$ is called a component of the link $L$. If we equip these knots with framing then we get a framed link. Framing of a knot $K$ is a normal vector field along $K$. If we push the knot $K$ along this vector field (slightly, such that it does not intersects with $K$) we get another knot $K'$. This defines a \emph{ribbon} bounded by $K$ and $K'$ -- see for instance Fig. 3 of \cite{Witten:1988hf}. So a framed knot is nothing but an orientable ribbon. The integer number counting the number of right-handed twists in the ribbon, or equivalently the linking number of $K$ with $K'$, is called the framing coefficient of the knot. For example, for $K$ the unknot with framing coefficient $n$, we have 
\[\begin{tikzpicture}[baseline=7,scale=2]
\draw [ thick]  (0,0,0) circle (.5);
\node[right]  at (-0.5,0,0) {$K^n$};
\end{tikzpicture}
\hspace{0.2 in} = \hspace{0.2 in}
\begin{tikzpicture}[baseline=7,scale=2]
\draw [ thick]  (0,0) circle (.5);
  \draw[thick,red] (0.388909,0.388909) to[out=135,in=45] (-0.388909,0.388909);
    \draw[thick,red] (-0.388909,0.388909) to[out=225,in=90] (-0.55,0);
      \draw[thick,red] (-0.55,0) to[out=-90,in=180] (0,-0.55);
     \draw[thick,red]  (0,-0.55)to [out=0, in = 225](0.388909,-0.388909);
      \draw[thick,red] (0.388909,0.388909) to[out=135,in=45] (-0.388909,0.388909);
      \draw[thick,red] (0.388909,0.388909) to [out=-45,in=210] (0.415692,0.24);
    \draw[thick,red]  (0.450333,0.26)  to[out=30,in=195]  (0.463644,0.124233);
    \draw[thick,red] (0.502281,0.134586) to[out=15,in=180] (0.48,0);
    \draw[thick,red] (0.52,0) to[out=0,in=170] (0.463644,-0.124233);
     \draw[thick,red](0.502281,-0.134586) to[out=-15,in=150] (0.415692,-0.24) ;
      \draw[thick,red](0.450333,-0.26) to[out=-30,in=225] (0.388909,-0.388909) ;
\node[left]  at (-0.55,0,0) {$\color{red}K'$};
\node[right]  at (-0.5,0,0) {$K$};
\node[right]  at (0.55,0,0) {$\Bigg \}$ $n$ times};
\end{tikzpicture}
\]

Consider a generic framed link, or equivalently ribbon graph, in $S^3$
\begin{equation}
	L = K_1^{n_1} \cup K_2^{n_2} \cup \cdots \cup K_m^{n_m}~, \label{link}
\end{equation}
where $K_i$ is the $i$-th component of the link with framing coefficient $n_i \in \bZ$. Given such a framed link in $S^3$, the following integral surgery procedure gives a closed and orientable 3-manifold:
\begin{enumerate}
	\item Drill out a tubular neighborhood $N_i$ (homeomorphic to the solid torus) of the knot $K_i$.
	\item Glue back $N_i$ with a homemorphism $h_i : \partial N_i \to \partial N_i \subset S^3$ such that
	\begin{equation}
		h_i : \mu_i \mapsto K_i' ~,\label{homeo}
	\end{equation}
	where $\mu_i$ is the contractible meridian curve of $N_i$ and $K_i' \subset \partial N_i$ is the knot specified by the framing of $K_i$ as explained above. In particular, the curve $K_i'$ becomes contractible after the gluing.
\end{enumerate}

All closed and orientable 3-manifolds can be obtained by such integral surgery procedure, and the resulting 3-manifolds depend only on the data specified by the framed link in equation \eqref{link}. However, there is a slightly more general procedure known as rational surgery if we allow the framing coefficients (\textit{a.k.a.} surgery coefficients) to be rational numbers instead of integers.

For rational surgery, we allow $h_i$ in \eqref{homeo} to be a homeomorphism specified by an arbitrary element of $SL(2,\bZ)$ (the mapping class group of $N_i$). To parameterize such elements, we pick a canonical basis for $H_1(\partial N_i)$. Take $\mu_i$ to be the meridian (hence trivial in $H_1(N_i)$), and $\lambda_i$ to be the longitude curve on the boundary of $N_i$. Then for any $h_i$ we can write
\begin{equation}
	h_i ([\mu_i]) = p_i [\mu_i] + q_i [\lambda_i] \in H_1(\partial N_i)  \,,
\end{equation}
where $[\mu_i]$ and $[\lambda_i]$ are the homology classes of the meridian and longitude, and $p_i, q_i \in \bZ$. The rational number $r_i = p_i/q_i \in \bQ$ is called the surgery coefficient of the knot $K_i$, and specifies the homology class of the curve $K_i'$ up to a sign.\footnote{Note that this does not determine the isotopy type of $h_i$ uniquely. However, the homeomorphism type of the resulting 3-manifold only depends on the link $L$ and the surgery coefficients.} This extends the surgery procedure explained above to the case where the surgery coefficients $n_i$ are rational numbers.

Note that the procedure of constructing 3-manifolds starting from a framed link is not unique. A 3-manifold can have different surgery presentations. For the case of integral surgery, any two such surgery presentations are related by a Kirby move \cite{kirby1978calculus}, and for rational surgery there are extra moves such as Rolfsen's twist to be able to go between different presentations \cite{rolfsen1984rational} (see \cite{gompf19994} for more details).

\subsection*{RT Invariant}

Here we review the calculation of RT invariant on 3-manifolds using their surgery presentation; for more details see \cite{Turaev+2010, bakalov2001lectures}.

\paragraph{Colored Framed Links (Anyon Amplitudes):} Given a unitary modular tensor category $\cC$ and a framed link $L \subset S^3$, we can label/color each knot component of $L$ with a simple object of $\cC$ to get a $\cC$-colored framed link. Moreover, if we choose an orientation on each knot, we get an oriented colored link. Such an oriented colored link is equivalent to a configuration of anyons linking with each other. By choosing a time direction on $S^3$, such a configuration describes a process in which anyons-antianyons are created from the vacuum and annihilate at a later time. Physically this process has an amplitude which we denote by $\langle L_\mathrm{col} \rangle \in \bC$. Since this process happens in a topological theory, the amplitude is a topological invariant and only depends on the isotopy type of the oriented colored link $L_\mathrm{col}$.

Using the braiding and modular matrices of $\cC$, the configuration of anyons described by $L_\mathrm{col}$ can be simplified to eventually calculate the amplitude $\langle L_\mathrm{col} \rangle \in \bC$ (\textit{cf.} equations \eqref{tspin} and \eqref{eq:defofS}). 
Having defined the anyon amplitudes, we can now define the RT invariant associated with $\cC$.  Take a 3-manifold $M_3$ that can be obtained from integral surgery on the framed link
\begin{equation}
	L = K_1^{n_1} \cup K_2^{n_2} \cup \cdots \cup K_m^{n_m} \label{link2}\, .
\end{equation}
The RT-invariant is then given by \cite{Turaev+2010}
\begin{equation}
	\mathrm{RT}(\cC,M_3) = \frac{e^{-2\pi i \sigma(L) c_- / 8}}{\cD^{m+1}} \sum_{a_1, \dots, a_m \,\in\, \cal I} d_{a_1} d_{a_2} \cdots d_{a_m} \, \big \langle L(a_1,\dots,a_m) \big \rangle  \,, \label{RT}
\end{equation}
where $\cal I$ is the set of simple anyons in $\cC$, $L(a_1,\dots,a_m)$ is the anyon configuration described by inserting anyon $a_i$ with framing $n_i$ on the knot component $K_i$ of $L$, $\cD = \sqrt{\mathrm{dim}(\cC)}$, and $\sigma(L)$ is the signature of the linking matrix of $L$.\footnote{To evaluate \eqref{RT} we need to choose an orientation on $L$, but it is clear that the final answer does not depend on the choice of orientation since we are summing over all anyons.}

For the case of manifolds obtained by plumbing on trees, the RT invariant can be expressed using only the modular $S$ and $T$ matrices \cite{Freed:1991wd}. In particular, for the lens space $L(p,q)=S^3/\bZ_p$ partition functions, we have
\begin{equation}
	Z\big[ L(p,q) \big] = (ST^{a_1}S\cdots S T^{a_n} S)_{00}~, \label{lens}
\end{equation}
where
\begin{equation}
	\cfrac{p}{q} = a_1 - \cfrac{1}{a_2-\cfrac{1}{ \cdots -\cfrac{1}{a_{n}} }}~.
\end{equation}

\section{Partition Function of Abelian CS Theories \label{app.Abelian}}

\subsection*{The $(E_8)_1$ Theory}

The $(E_8)_1$ theory is a trivial massive theory with no nontrivial anyons which explicitly allows to shift the infrared value of $c_-$ by 0 mod 8. 
The $(E_8)_{\pm1}$ theory is therefore an invertible field theory  with $c_-=\pm 8$. 
Stacking this theory, we can therefore generate all the possible invertible field theories with $c_-=0$ mod 8.

The $(E_8)_1$ CS theory can be constructed from the following K-matrix 
\begin{equation}\label{Eeight}
	{\bf K} =\left(\begin{matrix} 2 & -1 &0&0&0&0&0&0 \\ -1&2&-1&0&0&0&0&0 \\0&-1&2&-1&0&0&0&0 \\ 0&0&-1&2&-1&0&0&0 \\0&0&0&-1&2&-1&0&-1 \\0&0&0&0&-1&2&-1&0 \\0&0&0&0&0&-1&2&0 \\0&0&0&0&-1&0&0&2  \end{matrix}\right)~,
\end{equation}
which is the Cartan matrix of $E_8$ Lie algebra. It is straightforward to verify that there is only a trivial anyon since $\det{\bf K}=1$, and also that ${\rm sgn}({\bf K})=8$.

\subsection*{Coprime Matrices}

 In Section \ref{sec:anotherpov} and the remainder of this appendix, we will make use of various properties of coprime matrices that we now introduce. Two $n\times n$  integral matrices $C$ and $D$ (both assumed to have non-zero determinant) are said to be coprime if there exist integral matrices $A$ and $B$ such that 
\begin{equation}
	\label{Bezout.app}-BC^\mathrm{T} +AD^\mathrm{T} =\mathbbm{1}~.
\end{equation}
(The particular signs and transposition above are purely for later convenience.) 
We will find the following theorem useful~\cite{vaidyanathan2011general},
\begin{thm}
 Consider the ${2n \choose n}$ $n\times n$ minors of the matrix $(C~D)$. Then $C$ and $D$ are left coprime if and only if the $\mathrm{gcd}$ of all these minors is equal to unity.
\end{thm}
\noindent
In particular, this theorem implies that if $\mathrm{gcd}(\det C,\det D)=1$, then the matrices are coprime. 

An additional useful criterion is that if $DC^\mathrm{T}$ is an even symmetric matrix, i.e.\ $DC^\mathrm{T}=CD^\mathrm{T}$ and the integers on the diagonal of $DC^\mathrm{T}$ are all even, then one can make a ``preferred'' choice of $A$ and $B$ in~\eqref{Bezout.app} such that we also have $AB^\mathrm{T}=BA^\mathrm{T}$ and thus $\left(\begin{matrix}A & B \\ C & D\end{matrix}\right)$ is symplectic \cite{maass1954lectures}. In this case the inverse matrices are symplectic and integral as well, and hence we have additional relations
\begin{equation}
	A^\mathrm{T}D-C^\mathrm{T}B=\mathbbm{1}~,  \quad A^\mathrm{T}C=C^\mathrm{T}A~,\quad B^\mathrm{T}D=D^\mathrm{T}B~.
\end{equation}

In the context we will be interested in, we will have $D=\bfK\otimes \mathbbm{1}$ and $C=\mathbbm{1}\otimes \bfL$. These matrices commute and each of them is symmetric, and thus they form a symmetric pair. This symmetric pair is even because the elements on the diagonal of $\bfK$ are even. If we assume $\mathrm{gcd}(\det \bfK,\det \bfL)=1$ then we immediately infer that $\mathrm{gcd}(\det (\bfK\otimes \mathbbm{1}), \det (\mathbbm{1}\otimes \bfL))=1$ and hence these $C$ and $D$ are coprime. We now ask if the converse also holds: namely, does coprimality of $\bfK\otimes \mathbbm{1}$ and $\mathbbm{1}\otimes \bfL$ imply that $\mathrm{gcd}(|\det \bfK|, |\det \bfL|)=1$? 

The answer is in fact yes, as was quoted in Section \ref{sec:anotherpov}. To show this, assume to the contrary that $\mathrm{gcd}(|\det \bfK|, |\det \bfL|) \neq 1$ . Then there should exist a common prime factor $p$ that divides both $\det \bfK$ and $\det \bfL$. By Cauchy's theorem, there then exist elements (anyons) in Abelian groups $\mathbb{Z}^{|\bfK|}/\bfK\cdot \mathbb{Z}^{|\bfK|}$ and $\mathbb{Z}^{|\bfL|}/\bfL\cdot \mathbb{Z}^{|\bfL|}$ whose order are equal to $p$. If we represent such elements by integer vectors $\bf v$ and $\bf w$ respectively, we have
\begin{equation}
	p \,{\bf v} = \bfK \cdot {\bf x} \,, \qquad  p\, {\bf w} = \bfL \cdot {\bf y}
\end{equation}
for integer vectors $\bf x$ and $\bf y$ such that the vectors $\frac1p \bf x$ and $\frac1p \bf y$ are not integer. Consider now the vector $\frac1p {\bf x} \otimes \bf y$. The action of both $\bfK\otimes \mathbbm{1}$ and $\mathbbm{1}\otimes \bfL$ on this vector produces an integral vector,
\begin{equation}
	\bfK\otimes \mathbbm{1} \cdot \left( \frac1p {\bf x} \otimes {\bf y} \right) = {\bf v} \otimes {\bf y} \,, \qquad  \mathbbm{1}\otimes \bfL \cdot \left( \frac1p {\bf x} \otimes {\bf y} \right) = {\bf x} \otimes {\bf w} \,.
\end{equation}
But if the matrices $\bfK\otimes \mathbbm{1}$ and $\mathbbm{1}\otimes \bfL$ were coprime, then by $\eqref{Bezout.app}$ we would have concluded that the vector $\frac1p {\bf x} \otimes {\bf y}$ must be integer. But this is impossible because if $\frac1p {\bf x} \otimes \bf y$ is integer, $p$ should divide $x_i y_j$ for all $i$ and $j$. So either all $x_i$ or all $y_j$ are integer, which contradicts our assumption. Therefore if $\mathrm{gcd}(|\det \bfK|, |\det \bfL|) \neq 1$ the matrices $\bfK\otimes \mathbbm{1}$ and $\mathbbm{1}\otimes \bfL$ cannot be coprime.

\subsection*{Symmetry Between the Chern-Simons and Surgery Matrix}

In (\ref{eq:ZKLident}) of the Introduction, we quoted an identity relating the partition function of the Abelian TQFT with K-matrix $\bfK$ on a 3-manifold with linking matrix $\bfL$ to the same quantity with $\bfK$ and $\bfL$ exchanged. In this final subsection we  derive this identity. 

To begin, recall from Appendix \ref{app:RT} that in the surgery presentation the partition function can be obtained by summing over all anyons running on all knot components of the link:
\begin{equation}\label{partfun.app}
	Z_{\bfK}[\bfL]={1\over |\det \bfK|^{|\bfL|/2+1/2}}\sum_{{\bf m}\in {\mathbb{Z}}^{|\bfK||\bfL|}/ ({\bfK}\otimes \mathbbm{1}) \cdot {\mathbb{Z}}^{|\bfK||\bfL|} }e^{\pi i \, {\bf m}^\mathrm{T} ({\bfK}^{-1} \otimes {\bfL}) {\bf m}}~.
\end{equation}

For coprime matrices $C$ and $D$ such that $D C^\mathrm{T}$ is even and symmetric, there is an identity generalizing Gaussian reciprocity ~\cite[Theorem 1]{styer1984evaluating}:
\begin{equation}\label{Sthm}
	{1\over \sqrt{|\det D|}}\sum_{{\bf m}\in {\mathbb{Z}}^{|D|}/D^\mathrm{T}\cdot {\mathbb{Z}}^{|D|} }e^{\pi i \, {\bf m}^\mathrm{T} (D^{-1}C) {\bf m}}={1\over \sqrt{|\det C|}}e^{{2\pi i\over 8} {\rm sgn}({DC}^\mathrm{T})}\sum_{{\bf m}\in {\mathbb{Z}}^{|C|}/{C^\mathrm{T}}\cdot {\mathbb{Z}}^{|C|} }e^{- \pi i \, {\bf m}^\mathrm{T} (C^{-1}D) {\bf m}}~.
\end{equation}
The reciprocity law says that we can exchange $C$ and $D$ up to an 8th root of unity and complex conjugation.

Choosing $D=\bfK\otimes \mathbbm{1}$ and $C=\mathbbm{1}\otimes \bfL$, the LHS of \eqref{Sthm} reduces to \eqref{partfun.app} up to a factor of $\sqrt{|\det \bfK|}$. This immediately tells us that the theory with CS matrix $\bfK$ on a 3-manifold with linking matrix $\bfL$ gives, up to some normalization factors, the same result as the theory on linking matrix $\bfK$ with CS matrix $\bfL$.   

The theory with CS matrix $\bfL$ may be spin however. So for the proper physical interpretation of the statement that $\bfL$ and $\bfK$ are interchangeable, we need to discuss the partition functions of spin TQFTs. The basic issue here is that in a spin theory the topological spin
\ie
\theta({\bf m}) = e^{\pi i \, {\bf m}^\mathrm{T} \bfK^{-1}{ \bf m}}
\fe
is not a well-defined function on the quotient ${\bf m}\simeq {\bf m}+\bfK\cdot {\bf p}$ for integer vectors $\bf p$. This is because in a spin theory, there exists a transparent fermion which has spin $\frac{1}{2}$ and braids trivially with all other anyons.\footnote{For a discussion of the spin of line defects in spin vs. non-spin theories see for instance~\cite{Hsin:2019gvb,Ang:2019txy}.}

However, the braiding of the anyons given by
\ie
 B({\bf m}, {\bf n}) = e^{2\pi i \, {\bf m}^\mathrm{T} \bfK^{-1} \bf n}
 \fe
is still well-defined. Moreover by choosing a so-called integral Wu class $\bf W$ we can define a quadratic refinement of the braiding bilinear form as~\cite{Belov:2005ze}
\begin{equation}
	q({\bf m}) = e^{\pi i \,  ({\bf m}-\frac12 {\bf W})^\mathrm{T} \bfK^{-1} ({\bf m}-\frac12 {\bf W})}~, \label{q}
\end{equation}
which satisfies
\begin{equation}
	q({\bf m} + {\bf n}) - q({\bf m}) - q({\bf n}) + q(0) = B({\bf m}, {\bf n})~.
\end{equation}
The defining property of $\bf W$ is that ${\bf W}^\mathrm{T} \cdot {\bf p} = {\bf p}^\mathrm{T} \cdot \bfK \cdot {\bf p}$ mod 2 for all integral vectors $\bf p$. This means in particular that $W_i = K_{ii}$ mod 2 for all $i=1,\dots, |\bfK|$. With this we find that $q$ is a well-defined function on the quotient $\bZ^{|\bfK|}/\bfK\cdot \bZ^{|\bfK|}$ defining the anyons.

Note that the defining property of $\bf W$ only fixes it up to shifts by $2{\bf p}$ where $\bf p$ is any integral vector. We must thus understand how this affects the quadratic function $q$. For this we note that 
\ie
q({\bf m}) = \theta\Big({\bf m} - \frac{1}{2} {\bf W}\Big)~.
\fe
From this we see that shifting $\bf W$ by $2\bf p$ amounts to a permutation of the anyons labels via the redefinition ${\bf m} \mapsto {\bf m}-{\bf p}$.

As pointed out in \cite{Belov:2005ze}, the crucial difference between the quadratic refinement $q$ in spin theories and $\theta$ in non-spin theories is the fact that $\theta(0)=1$, whereas in general $q(0)\neq1$. For this reason we cannot think of $q$ as the spin of anyons. However, when there is a choice of $\bf W$ such that $q(0)=1$, the theory can be viewed as a non-spin theory. This is because in that case the transparent fermion decouples from the detectable anyons, and the theory becomes equivalent to the tensor product of a bosonic theory with $\theta=q$ and the trivial spin theory containing the transparent fermion.

Similar to the bosonic case, we can compute the chiral central charge, or equivalently the $L(1,1) \simeq S^3$ partition function, as~\cite{Belov:2005ze}
\begin{equation}
	e^{2\pi i \frac{c_-}{8}} = \frac{1}{\sqrt{|\det \bfK|}} \sum_{{\bf m}\,\in\, {\mathbb{Z}}^{|\bfK|}/{\bf K}\cdot {\mathbb{Z}}^{|\bfK|}}  e^{\pi i\, ({\bf m}-\frac12 {\bf W})^\mathrm{T} {\bf K}^{-1} ({\bf m}-\frac12 {\bf W})}~.
\end{equation}
Therefore $\bf W$ enables us to compute the $S^3$ partition function which admits a unique spin structure. We now turn to the situation that there is a linking matrix $\bfL$. In this case we must guess what the appropriate quantity to sum over is. The natural guess is
\begin{equation}
	Z_{\bfK}[\bfL,x]={1\over |\det \bfK|^{|\bfL|/2+1/2}}\sum_{{\bf m}\in {\mathbb{Z}}^{|\bfK||\bfL|}/ ({\bfK}\otimes \mathbbm{1}) \cdot {\mathbb{Z}}^{|\bfK||\bfL|} }e^{\pi i \, ({\bf m}-\frac{1}{2} {\bf W} \otimes {\bf x})^\mathrm{T} ({\bfK}^{-1} \otimes {\bfL}) ({\bf m}-\frac{1}{2} {\bf W} \otimes {\bf x})}~.
\end{equation}
But for this to be well-defined under the usual shifts ${\bf m}\to {\bf m}+\bfK \cdot {\bf p}$ for arbitrary integral ${\bf p} \in \bZ^{|\bfK||\bfL|}$, we need that 
\begin{equation}
	({\bf W} \otimes {\bf x})^\mathrm{T} (\mathbbm{1} \otimes \bfL) \cdot {\bf p} = {\bf p}^\mathrm{T} \cdot(\bfK \otimes \bfL)\cdot {\bf p} \pmod{2} \label{wx}
\end{equation}
holds for all $\bf p$. This forces us to choose $L_{AB} W_{i} x_B = K_{ii} L_{AA}$ mod 2 for all $i=1,\dots,|\bfK|$ and $A,B=1,\dots,|\bfL|$. Since $W_i = K_{ii}$ mod 2, this is equivalent to $L_{AB} x_B = L_{AA}$ mod 2 (assuming $\bfK$ is odd). One obvious freedom in this equation is to shift ${\bf x} \to {\bf x} + 2 {\bf s}$ for any integral vector ${\bf s} \in \bZ^{|\bfL|}$. As before, this can be reabsorbed by redefining the anyons. Therefore the space of inequivalent solutions is
\ie
\left\{{\bf x} \in \bZ_2^{|\bfL|} :  \bfL \cdot {\bf x} = {\bf x} \mod{2} \right\}~.
\fe
These solutions correspond to what are known as \textit{characteristic sublinks} of $\bfL$, and are shown to be in one-to-one correspondence with the spin structures on the 3-manifold associated with $\bfL$~\cite[Appendix C]{kirby19913} (see also \cite{blanchet1992invariants}).  This shows that spin theories require a choice of spin structure in order for the partition function to be well defined, as one would expect.

However, when $\bfL$ is even (or more generally when $\bfK \otimes \bfL$ is even), ${\bf W} \otimes {\bf x} = 0$ is always an allowed solution of \eqref{wx}. In other words when the surgery linking matrix is even, there is a preferred spin structure ${\bf x}=0$ given by the particular surgery presentation $\bfL$. Hence we can now interpret of the sum \eqref{partfun.app} for when $\bfK$ is odd but $\bfL$ is even as the partition function of the Abelian spin CS theory $\bfK$ on 3-manifold $\bfL$ with some specific spin structure. Using the reciprocity formula \eqref{Sthm}, and following the normalization of partition functions we find the aforementioned symmetry:
\begin{equation}
	 \sqrt{|\det \bfK|} \, Z_{\bfK}[\bfL] =e^{{2\pi i \over 8}\mathrm{sign}(\bfK)\mathrm{sign}(\bfL)} \sqrt{|\det \bfL|} ~\overline {Z_\bfL[\bfK]}~. \label{K.L.symmetry}
\end{equation}

One could imagine a tentative explanation of this symmetry in terms of the 3d-3d correspondence~\cite{Dimofte:2011ju} as follows. Both sides of \eqref{K.L.symmetry} can be perhaps interpreted as the partition function of the 6d $\mathcal{N}=(2,0)$ theory of $U(1)$-type considered in \cite[equations (2.23)-(2.25)]{Gadde:2013sca} on the 6-manifold given by the tensor product of 3-manifolds with linking matrices $\bfK$ and $\bfL$.
We can compactify the 6d theory on either of the two 3-manifolds. In each case, after the compactification we find a 3d $\mathcal{N}=2$ pure CS theory with K-matrix $\bfK$ or $\bfL$. In such supersymmetric theories, the fermions decouple and after integrating them out contribute some extra normalization factor. 
It would be interesting to confirm that integrating out these fermions gives the same normalization factor in \eqref{K.L.symmetry}.

\bibliography{refs}
\bibliographystyle{JHEP}

\end{document}